\documentclass[pra,twocolumn,showpacs,amsmath,amssymb,superscriptaddress,floatfix]{revtex4-1}
\usepackage{graphicx,color}
\usepackage{amsmath,amssymb,bm}
\usepackage{braket}	
\usepackage{epsfig}

\definecolor{darkred}{rgb}{0.90,0,0}
\definecolor{darkgreen}{rgb}{0,0.60,.2}
\definecolor{darkblue}{rgb}{0,0,1}
\definecolor{grey}{cmyk}{0,0,0,0.25}
\definecolor{orange}{cmyk}{0,0.6,0.8,0}

\begin{document}
\title{Interaction quantum quenches in the one-dimensional Fermi-Hubbard model with spin imbalance}

\author{L. Riegger}
\affiliation{Department of Physics and Arnold Sommerfeld Center for Theoretical Physics,
Ludwig-Maximilians-Universit\"at M\"unchen, D-80333 M\"unchen, Germany}
\affiliation{Max-Planck-Institute of Quantum Optics, D-85748 Garching, Germany}
\author{G. Orso}
\affiliation{Laboratoire Mat\' eriaux et Ph\' enom\`enes Quantiques,
Universit\' e Paris Diderot-Paris 7 and CNRS, UMR 7162, 75205 Paris Cedex 13, France}
\author{F. Heidrich-Meisner}
\affiliation{Department of Physics and Arnold Sommerfeld Center for Theoretical Physics,
Ludwig-Maximilians-Universit\"at M\"unchen, D-80333 M\"unchen, Germany}

\begin{abstract}
Using the time-dependent density matrix renormalization-group method and exact diagonalization, we study the nonequilibrium dynamics of the 
one-dimensional Fermi-Hubbard model following a quantum quench or a ramp of the on-site interaction strength.
We are particularly interested in the nonequilibrium evolution of Fulde-Ferrell-Larkin-Ovchinnikov (FFLO) correlations, which, at finite spin polarizations and for attractive interactions,
are the dominant two-body correlations in the ground state. For quenches from the noninteracting to the attractive regime, we investigate the dynamical emergence of FFLO correlations and their signatures in the pair quasi--momentum distribution function.
We observe that the
postquench double occupancy  exhibits a  maximum as the interaction strength becomes of 
the order of the bandwidth.
Finally, we study quenches and ramps from attractive to repulsive
interactions, which imprint FFLO correlations onto repulsively bound pairs.
 We show that a quite short ramp time is sufficient to wipe out the characteristic FFLO features in the 
postquench pair momentum distribution functions.
\end{abstract}

\maketitle


\section{Introduction}
\label{sec:intro}
The physics of low-dimensional Fermi gases with attractive interactions and unequal spin populations has generated 
interest because of the possibility of realizing exotic fermionic superfluids such as the 
Sarma \cite{sarma63} and Fulde-Ferrell-Larkin-Ovchinnikov (FFLO) states \cite{fulde64,larkin64} (see Refs.~\cite{sheehy07,chevy10,gubbels13} for a review).
The FFLO state is characterized by a spatially oscillating
  order parameter with the excess fermions sitting mainly in
   its nodes, where they are less detrimental to superconductivity. 
The FFLO state is currently invoked to explain the behavior of several superconducting systems, ranging from heavy fermion \cite{rado03} and organic \cite{uji06} materials to dense quark matter in the core of neutron stars \cite{casal04}. However, the experimental evidence of this phase in solid-state superconductors is still controversial.
In particular,
the one-dimensional (1D) analog of the FFLO state, a state with spatially oscillating
pairing correlations  is, at a finite spin imbalance,  the ground state of the Fermi-Hubbard
model and its continuum analog, the Yang-Gaudin model
(see Refs.~\cite{feiguin11} and \cite{guan13} for a review). A recent experiment \cite{liao10}
has realized a spin-imbalanced Fermi gas in an array of 1D systems and has demonstrated that the observed density
profiles are in agreement with theoretical predictions \cite{orso07,hu07,kakashvili09}. 
An actual experimental proof of the presence of a spatially modulated quasicondensate in this
system, expected from theory \cite{yang01,feiguin07,batrouni08,tezuka08,casula08,luescher08}, is still
lacking, which has led to a large number of proposals for 
schemes to probe FFLO correlations \cite{bakhtiari08,roscilde09,edge09,kajala11a,bolech12,lu12}.
Many of these proposals involve the presence of an optical lattice along the 1D direction,
suggesting that the Fermi-Hubbard model is the appropriate model. 
This system has been realized with ultracold atoms by several groups \cite{schneider08,joerdens08,hart14}.

One-dimensional gases also provide a natural arena for the study of the nonequilibrium dynamics of closed, strongly correlated many-body systems and several experiments were specifically tailored to explore this physics \cite{kinoshita06,hofferberth07,kasztelan12,trotzky12,gring12,cheneau12,ronzheimer13,langen13}. For instance, theoretical and experimental research is aiming at 
understanding the conditions for the emergence of thermalization \cite{rigol08,polkovnikov11} 
and the qualitative features in the relaxation dynamics such as  time scales for the approach to
the steady state (see Refs.~\cite{polkovnikov11,eisert14,langen14} for recent reviews). Integrable 1D 
models such as the Fermi-Hubbard model play a special role, since there, thermalization
can often only occur with respect to so-called generalized Gibbs ensembles \cite{rigol07}, a subject addressed
in a series of recent studies (see, e.g., \cite{essler12, caux12,Mierzejewski14,wouters14,pozsgay14,goldstein14,alba14}).

In this work we study the real-time dynamics of FFLO correlations and the double occupancy
in interaction quantum quenches in the Fermi-Hubbard model with spin imbalance. The Hamiltonian of the system
is given by
\begin{equation}
H= - J \sum_{i=1}^{L-1} (c_{i\sigma}^{\dagger}c_{i+1\sigma}+h.c.) + U \sum_{i=1}^L n_{i\uparrow}n_{i\downarrow}\,,
\label{eq:ham}
\end{equation}
where  $c_{i\sigma}$ annihilates a fermion with spin $\sigma=\uparrow,\downarrow$ at site $i$ of a chain of length $L$, $n_{i\sigma}=c^\dagger_{i\sigma} c_{i\sigma}$ is the density operator for the
 spin-$\sigma$ component, $U$ is the on-site interaction and $J$, the hopping parameter. 
Moreover, we define the total density, $n=(N_{\uparrow}+N_{\downarrow})/L$, and the spin polarization, $p=(N_{\uparrow}-N_{\downarrow})/(N_{\uparrow}+N_{\downarrow})$, $N_{\sigma}$ being the number of particles with spin $\sigma$.
The nonequilibrium dynamics in the Fermi-Hubbard model has  been studied in recent experiments \cite{strohmaier10,schneider12,will14,perthot14}.
Our main interest is in the relaxation dynamics, the identification of 
relevant time scales for the formation of FFLO correlations, and the dependence of 
observables on postquench parameters.

Based on two  numerical methods,
time evolution in a Krylov subspace  using exact diagonalization (ED; see, e.g., \cite{manmana05} and \cite{sandvik}) and
the time-dependent version of the density matrix renormalization-group (DMRG) technique \cite{vidal04,daley04,white04}, 
we investigate the behavior of the system following both instantaneous  and slow quenches of the interaction strength. 
In the latter case, we are interested in 
 the crossover
to the adiabatic regime  (see Refs.~\cite{bernier11} and \cite{bernier12} for studies of slow quenches in the Bose-Hubbard model). 
 
We first present results for quenches from the noninteracting case, $U=0$, to the attractive regime, $U<0$.  
In particular, we investigate how FFLO correlations emerge in the system 
and we  propose that a time scale can be extracted by monitoring the dynamics of natural orbitals, which are the eigenstates of the pair density matrix, and 
of the pair quasi--momentum distribution function (MDF). The latter quantity is a key observable
in the discussion of the FFLO state since the spatially inhomogeneous quasicondensate translates into 
maxima in the pair MDF at finite momenta $q$,  which  is  controlled via the polarization \cite{feiguin07,feiguin11} 
\begin{equation}
q=\pi n p \,.\label{eq:qnp} 
\end{equation}
Moreover, we analyze the relaxation dynamics of the double occupancy, a much discussed quantity in studies of 
quantum quenches in the repulsive Fermi-Hubbard model \cite{moeckel08,kollar08,eckstein09,eckstein10,eckstein11,hamerla13,hamerla14,hm09,kessler13,kajala11}, which has been measured in nonequilibrium experiments \cite{joerdens08,strohmaier10,ronzheimer13}.

We then turn our attention to quenches from the attractive regime, $U<0$, to the repulsive one, $U>0$. 
In particular, we demonstrate how FFLO correlations of Cooper pairs
can be imprinted  onto repulsively bound pairs. We find that the visibility of the FFLO peak
crucially depends on the final (postquench) value of $U$ and on the ramp time.

The plan of this paper is as follows. In Sec.~\ref{sec:setup}, we introduce  the quench schemes and provide definitions
of quantities used throughout the paper  and  details on our
numerical methods.
Section~\ref{sec:formfflo} is devoted to the formation of FFLO correlations 
in quenches and ramps starting from the noninteracting case to negative values of $U$.
In Sec.~\ref{sec:imprint}, we discuss the imprinting of FFLO correlations onto
repulsively bound pairs in quenches and ramps from $U<0$ to $U>0$. 
In Sec.~\ref{sec:sum}, we summarize our main results and discuss perspectives for future work on this exciting topic.

\section{Set-up, definitions, and methods}
\label{sec:setup}
\subsection{Quench schemes}
We consider two quench schemes. In the first case, 
we prepare the system in the ground state of Eq.~\eqref{eq:ham} at $U=0$, and at $t=0$ 
we change the interaction strength to $U<0$. This probes the formation of FFLO correlations
in a previously noninteracting two-component Fermi gas.
In the second scheme, the initial state is the ground state at some $U<0$, and then we instantaneously
change the sign of the interaction at $t=0$. This scheme is supposed to imprint FFLO 
correlations onto the repulsively bound pairs present on the repulsive side $U>0$.

Besides these quenches that change the value of $U$ abruptly from $U_i$ to $U_f$,
we also consider slow, linear quenches, referred to as ramps, where in the time interval $t \in \lbrack 0 , t_{\rm ramp} \rbrack$, $U$ takes  the time dependence
according to 
\begin{equation}
U(t) = U_i+ \frac{U_f-U_i}{t_{\rm ramp}} t\, , \label{eq:ramp} 
\end{equation}   
where $t_{\rm ramp}$ is the ramp time.

\subsection{Observables}
Our study mainly focuses on two  quantities, namely, the double occupancy and the pair MDF.
The double occupancy $d$  is defined as 
\begin{equation}
d = \sum_{i=1}^L \langle n_{i\uparrow} n_{i\downarrow} \rangle \,.
\end{equation}

The pair MDF $n_k^{\rm pair}$ (MDF) is the Fourier transform of pair correlations $\rho^{\rm pair}_{\ell j}$, where
\begin{equation}
\rho^{\rm pair}_{\ell j} = \langle c^{\dagger}_{\ell\uparrow}c^{\dagger}_{\ell\downarrow} c_{j\downarrow}c_{j\uparrow} \rangle
\label{eq:rhoij}
\end{equation}
such that 
\begin{equation}
n_k^{\rm pair} = \frac{1}{L} \sum_{\ell,j} e^{i (\ell -j ) k} \rho^{\rm pair}_{\ell j}\,.
\label{eq:pmdf}
\end{equation}
Note that the two quantities defined above are related to each other by the normalization condition $d=\sum_k n_k^{\rm pair}$.

While generally, there are established methods to measure the MDF of the fermionic components via time-of-flight and band-mapping techniques \cite{bloch08}, the measurement of the pair MDF in the FFLO phase has not been accomplished yet, but its observation is the goal of future experiments \cite{liao10}. 

We further define the visibility $V$ via
\begin{equation}
V= \frac{n_{k=q}^\mathrm{pair} - n_{k=0}^\mathrm{pair}}{n_{k=q}^\mathrm{pair} + n_{k=0}^\mathrm{pair}} \label{eq:visibility}\,.
\end{equation}
This particular definition is motivated by the fact that the maximum in $n_{k}^\mathrm{pair}$ is initially at $k=0$
for $U_i=0$ and at $k=\pm q$ in the ground state for  $U<0$.

\subsection{Numerical methods}
We employ two numerical methods in our work, namely, ED and the DMRG method \cite{vidal04,daley04,white04}. In the ED method, we use a Krylov-space method  to propagate the wave function in time (for technical details, see, e.g., \cite{manmana05}).
We use a Trotter-Suzuki method
to propagate the wave function using the DMRG; the time step is typically chosen to be $\delta t\sim 0.02/J$. The discarded
weight, which controls the accuracy of the truncation involved in the DMRG \cite{schollwoeck11,schollwoeck05}, is set to 
$10^{-6}$, which is sufficient to ensure negligible numerical errors for the times reached in the simulation.
We perform all DMRG and ED simulations for  open boundary conditions.

\section{Formation of FFLO correlations in quenches and linear ramps from $U=0$ to $U<0$}
\label{sec:formfflo}
In this section, we present our results for quenches and linear ramps from the noninteracting case
$U_i=0$ to attractive interactions $U_f<0$. 
First, we discuss the behavior of  the double occupancy and the pair MDF for the quantum quench 
in Sec.~\ref{sec:quench1} and then we  turn to slow quenches in Sec.~\ref{sec:ramps1}.

\subsection{Quantum quench}
\label{sec:quench1}

\begin{figure}[t]
\includegraphics[width=.96\columnwidth]{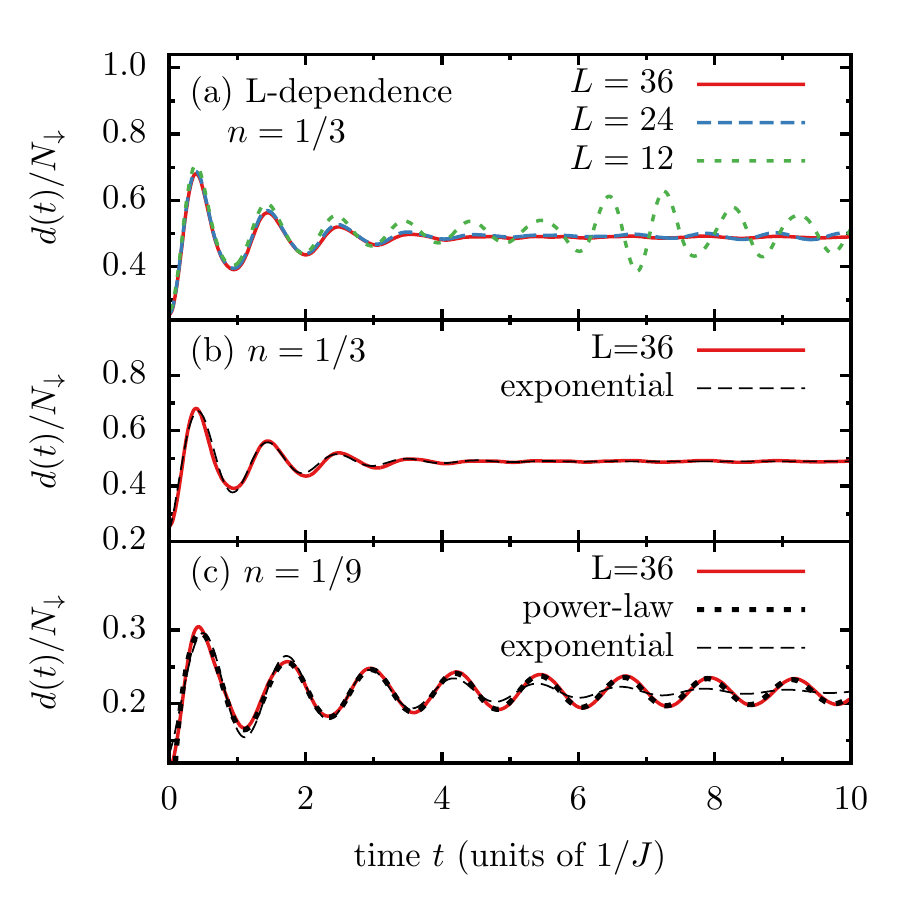}
\caption{(Color online) (a) System-size dependence [lattice sizes $L=12$ (ED), $L=24,36$ (DMRG)] of the time evolution of the  double occupancy $d(t)/N_{\downarrow}$ after the interaction quench from  $U_i=0$ to $U_f=-8J$ at $n=1/3$, $p=1/2$. (b) Fit (dashed black line) of a single, exponentially damped oscillation $-a\,\cos(\omega t+\phi)\exp(-t/\tau_d)+\bar{d}$ to the  dynamics of $d(t)$ [for $t\lesssim 3/J$: $\tau_d=(1.05\pm 0.02)/J$, $\omega=(6.15\pm 0.02)J$, $\bar{d}$ from $t>3/J$]. Note that we observe a short relaxation time , $\tau_d\sim  1/J$, for this set of parameters. 
(c) $d(t)$ at low densities. In this regime, the relaxation follows a power-law (dotted black line), $-a\,\cos(\omega t+\phi)/t^\alpha+\bar{d}$ [for $0.1/J\lesssim t\lesssim 5/J$: $\alpha=0.53\pm 0.01$, $\omega=(5.097\pm 0.005)J$] and cannot be described by the fit of an exponentially damped oscillation (dashed black line).}
\label{fig:do_dynamics}
\end{figure} 

\subsubsection{Double occupancy}
\label{sec:d_quench1}

{\it Example for time evolution. }
Typical results for the time dependence of the double occupancy are shown in Fig.~\ref{fig:do_dynamics}(a), where we display ED data for $L=12$ and DMRG data for $L=24,36$, $U_f=-8J$, polarization $p=1/2$, and filling $n=1/3$.
Since the system is initially prepared in an uncorrelated state, the double occupancy at $t=0$ is given by 
$d(0)=n_\uparrow n_\downarrow L$; that is,
\begin{equation}\label{din}
\frac{d(0)}{N_\downarrow}=\frac{n(1+p)}{2}.
\end{equation}
Similarly to the behavior of the double occupancy in other interaction quantum quenches \cite{ronzheimer13,sorg14}, $d(t)$ undergoes a fast transient characterized by large oscillations and relaxes to a constant value $\bar{d}$ on a 
time scale $\tau_d$ of order $1/J$,  as is evident from the data for $L=36$. For small systems such as $L=12$, there are 
revivals due to reflections from the boundary of the simulation box. These spurious oscillations quickly disappear
as $L$ increases.

\begin{figure}[t]
\includegraphics[width=.96\columnwidth]{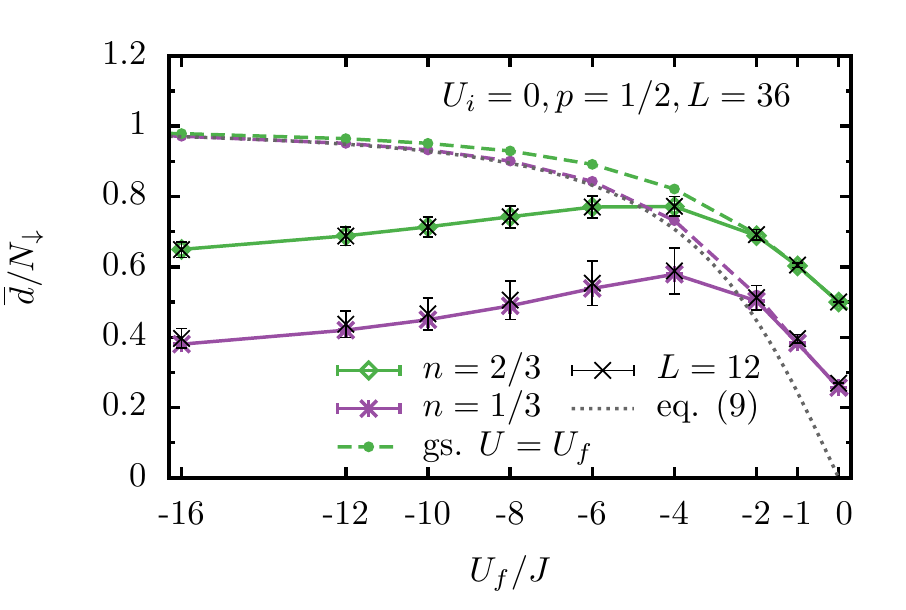}
\caption{(Color online) Time average $\bar d$ of the double occupancy $d(t)$ versus the postquench interaction strength $U_f$ after  the quench from $U_i=0$, for different fillings, $n=2/3$ (diamonds) and $n=1/3$ (asterisks), for $L=36$ (DMRG). Black X's indicate the corresponding results for $L=12$ (ED). Dashed lines show the ground-state values for $U=U_f$. Error bars indicate the standard deviation from the mean value (not visible for $L=36$). The dotted line shows the result for the low-density limit, Eq.~\eqref{eq:vanishing-density-limit}.
}\label{fig:do_avrg_vs_interaction}
\end{figure}

\begin{figure}

\includegraphics[width=.96\columnwidth]{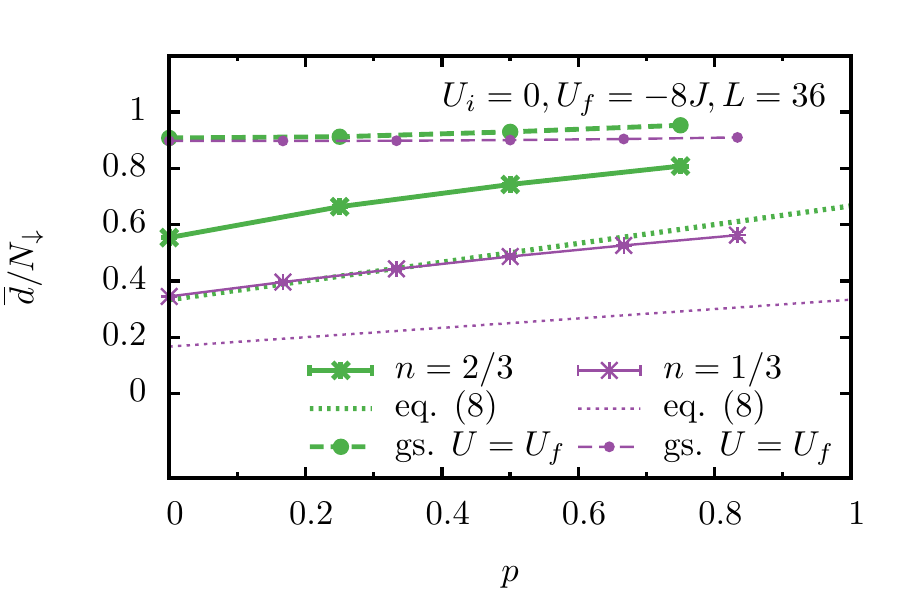}
\caption{(Color online) Time average $\bar d$ of the  double occupancy after  the quench from $U_i=0$ to $U_f=-8J$ as a function  of the  polarization $p$. Thin and thick lines correspond to fillings $n=1/3$ and $n=2/3$, respectively. Dotted lines show the prequench value of $\bar d$ from Eq.~\eqref{din};  dashed lines indicate the postquench ground-state value with $U=U_f$.
All data for $L=36$ (DMRG); the time average was computed for $t>3/J$.}\label{fig:do_avrg_vs_polarization}
\end{figure}

{\it Long-time average $\bar d$.}
We extract the stationary value $\bar d$ by averaging the double occupancy $d(t)$ at long times $(t \gtrsim 3/J)$.
These data are plotted in Fig.~\ref{fig:do_avrg_vs_interaction} as a function of the postquench interaction strength $U_f/J$ 
for two values of the filling, namely, $n=1/3$ and $n=2/3$, and polarization $p=1/2$. 
As $|U_f|/J$ increases, we see that initially $\bar d$  increases from its $U=0$ value until it reaches a maximum  
value as the interaction strength becomes of the order of the bandwidth, $|U_f|/J\sim 4$, and then it decreases (slowly)
for larger values of $|U_f|$.

It is instructive to compare the steady-state double occupancy $\bar d$ to its ground-state value  $d_{\rm gs}$, calculated
for $U=U_f$. The latter is shown in Fig.~\ref{fig:do_avrg_vs_interaction} by  dashed lines. 
Different from $\bar d$, the ground-state double occupancy is  a monotonically increasing function of $|U|/J$, approaching the largest possible value of $d_{\rm gs} = N_{\downarrow} $ for 
$|U|/J\to \infty$ (note that we plot the ratio $\bar d/N_{\downarrow}$ in the figure).
In particular, taking into account that $d_{\rm gs}=\partial E_\textrm{gs}/\partial U$, where
$E_\textrm{gs}=-\sqrt{U^2+16J^2}$ is the ground-state energy of the two-body problem,
 for vanishing density, $n\rightarrow 0$, we find (dotted line in Fig.~\ref{fig:do_avrg_vs_interaction})
\begin{equation}
\label{eq:vanishing-density-limit}
\frac{d_\textrm{gs}}{N_\downarrow}=\frac{-U}{\sqrt{U^2+16J^2}}.
\end{equation}

We see in Fig.~\ref{fig:do_avrg_vs_interaction}  that for sufficiently weak interactions, $|U_f|\ll J$, the steady-state double occupancy follows closely the postquench 
ground-state value,  while for stronger interactions  it is well below that value. This effect can be  understood by taking into account that during the time evolution the energy is conserved, i.e.,  
\begin{equation}\label{Econs}
E_\textrm{kin}(0)+U_f d(0)=E_\textrm{kin}(t)+U_f d(t),
\end{equation}
where $E_\textrm{kin}$ is the expectation value of the kinetic energy. Equation~(\ref{Econs}) shows that an increase in the
double occupancy has to be compensated by  a corresponding change in the kinetic energy. 
Since the single-particle dispersion has a finite bandwidth $4J$,  the conversion between interaction 
and kinetic energies is progressively inhibited for $|U_f|>4J$, implying that $\bar d$ decreases 
by further increasing the interaction strength.
In particular, for $U_f/J\rightarrow -\infty$ the double occupancy remains frozen to its initial value, $\bar d=d(t)=d(0)$, 
with the latter given by Eq.~(\ref{din}).

\begin{figure}[t]
\includegraphics[width=.96\columnwidth]{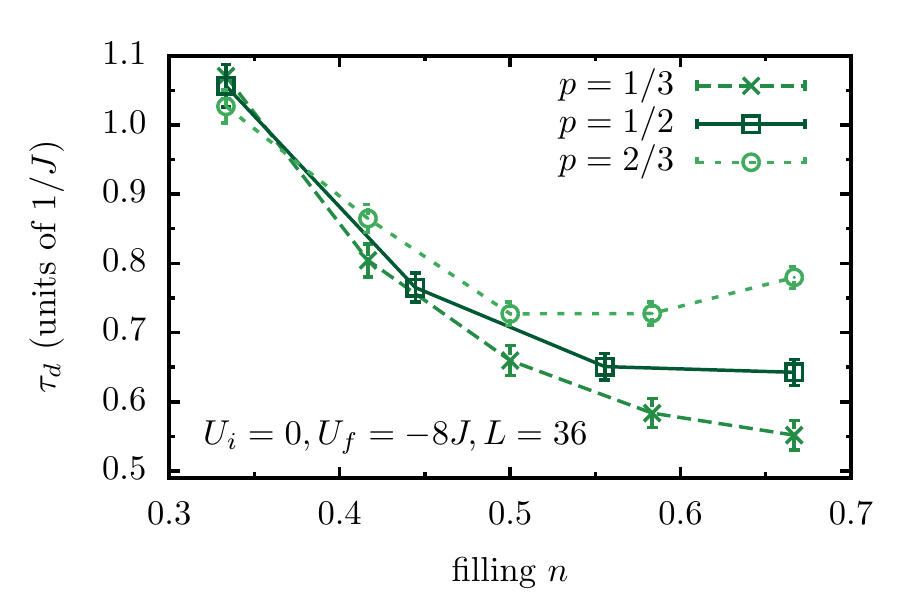}
\caption{(Color online) Relaxation time $\tau_d$ of the double occupancy in the quench from $U_i=0$ to $U_f<0$ versus
filling $n$ for different polarizations $p$ and $U_f=-8J$ ($L=36$; DMRG). 
Error bars indicate the asymptotic standard error from the least-squares fit of Eq.~\eqref{eq:fit_d} to the data.
}\label{fig:do_buildup}
\end{figure}

In Fig.~\ref{fig:do_avrg_vs_polarization}, we show the dependence of $\bar d$ on the spin polarization for fixed
interaction strength $U_f=-8J$ and for two fillings, $n=2/3$ (diamonds) and $n=1/3$ (asterisks). 
$\bar d$ increases roughly linearly as a function of $p$, as is the case for the initial double occupancy [see Eq.~(\ref{din})].

{\it Relaxation towards the stationary regime.} To further analyze the data, 
we observe that the fit function
\begin{equation} \label{eq:fit_d}
d(t)=-a\,\cos(\omega t+\phi)\exp(-t/\tau_d)+\bar{d}
\end{equation} 
provides a reasonably good description of the numerical data in a wide parameter regime. An example  is shown in Fig.~\ref{fig:do_dynamics}(b) for the largest system size, $L=36$.

The frequency of the coherent oscillation is, for large values of $|U_f|/J$, given by $\omega \propto |U_f|$, similar to other studies of collapse and revival phenomena \cite{greiner02,will14,iyer14},
since the dynamics is predominantly governed by the interaction term
in that limit. By the same argument, $\omega$ increases with density, since the double occupancy increases with density, both at  
equilibrium and in the steady state.
In the two-body limit ($N_\uparrow=N_{\downarrow}=1$), the frequency is given by the binding energy $\epsilon_b$:
\begin{equation}
\omega = \epsilon_b = \sqrt{U^2+16J^2}- 4J\,. 
\end{equation}

In particular, the good agreement between Eq.~\eqref{eq:fit_d} and the data shown in Fig.~\ref{fig:do_dynamics}(b) suggests  that there is an exponential decay towards the time-independent value $\bar{d}$, which allows us to extract 
the relaxation time $\tau_d$. 
 
We observe that $\tau_d$ varies only mildly with both polarization and postquench interaction strength $U_f$, whereas the dependence on $n$
is much stronger. 
This is illustrated in Fig.~\ref{fig:do_buildup}, where we plot $\tau_d$ versus $n$.
Starting at low densities, $\tau_d$ decreases  as $n$ increases and then becomes as short as $\tau_d \sim 0.5/J$. 
For large polarizations, however, $\tau_d$ increases again for $n\to 1$.
Several aspects play a role in the relaxation dynamics: (i) the presence of doublons defined by the initial state at $t=0$, (ii)
the formation of additional doublons and (iii) the scattering of doublons, which leads to the decay of $d(t)$ to the stationary value.
The density dependence can be understood from the simple argument that the higher the density, the higher is the probability of scattering events. 
At a high density, these events can occur on  times scales set by the inverse hopping rate, resulting in $\tau_d \sim 1/J$.

In  the opposite limit  of low densities $n\lesssim 0.2$, the decay  of the amplitude of the oscillations is actually better described by a power-law
than by an exponential. This is illustrated    in Fig.~\ref{fig:do_dynamics}(c), where the fits  to both a power law and an exponential relaxation are shown.
One can therefore attempt to describe the crossover from a power-law, realized at small $n$, to an exponential 
relaxation at large $n$  with a function 
\begin{equation}
d(t)=-a\,\cos(\omega t+\phi)  \exp(-t/\tau_d)/t^{\alpha}+\bar{d}\,.
\end{equation}
Therefore, the power-law decay and the coherent oscillations are inherited from the few-body dynamics, whereas the many-body physics introduces an actual damping.
Consistent with this picture, we observe that the exponent of the power-law $\alpha \approx 0.5 $ very weakly depends on $U_f/J$
and that $\tau_d$ becomes very large for $n\to 0$.

\begin{figure}[!t]
\includegraphics[width=.96\columnwidth]{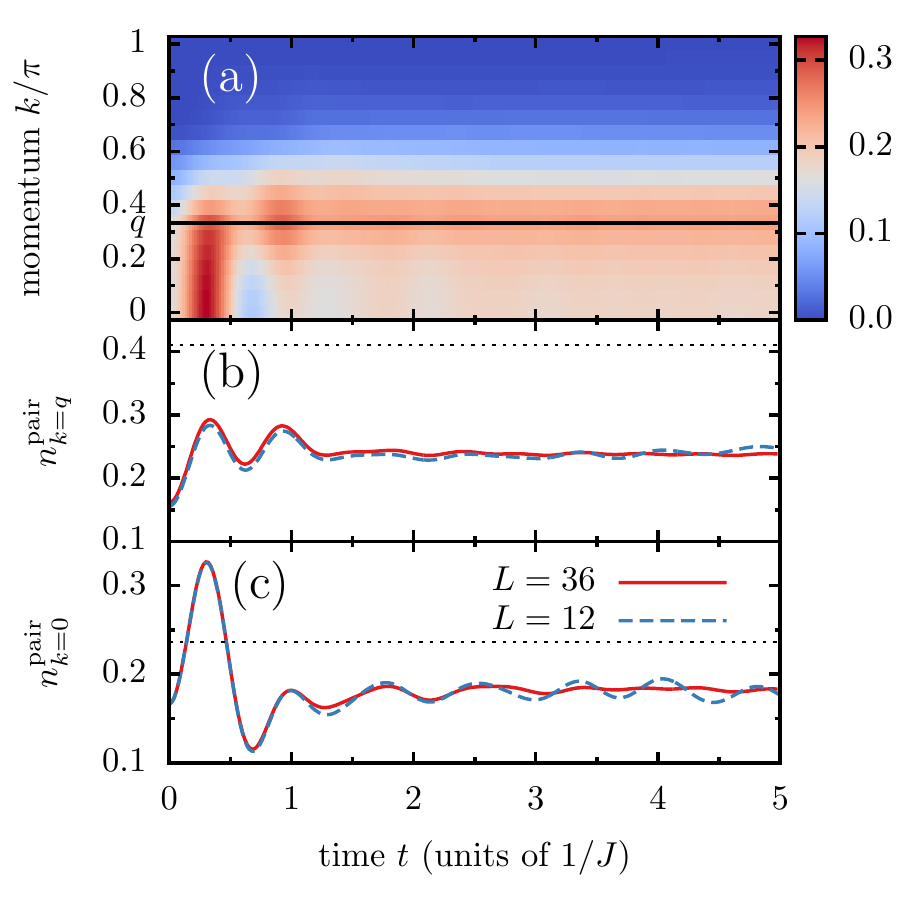}
\caption{(Color online) (a) Pair momentum distribution $n_{k}^\mathrm{pair}(t)$ after the interaction quench from  $U_i=0$ to $U_f=-8J$ at $n=1/3$, $p=1/2$, $L=36$ (DMRG); the horizontal line marks the position of the FFLO momentum $q$ set by the density and polarization. (b)
$n_{k}^\mathrm{pair}(t)$ at the FFLO momentum $q$.  (c)  $n_{k=0}^\mathrm{pair}(t)$ at zero momentum; solid lines for $L=36$ (DMRG) and dashed lines for $L=12$ (ED). Dotted horizontal lines in (b) and (c) indicate postquench ground-state values of the respective variables.
}\label{fig:pairMomDistr_vs_time}
\end{figure}

\subsubsection{Pair correlations and quasi--momentum distribution function}

{\it Time evolution of the pair MDF.}
We next turn to the main quantity of interest in our study, namely the pair MDF $n_k^{\rm pair}$. 
A typical example of its time evolution in the quench from $U_i=0$ to $U_f=-8J$  is shown in 
Fig.~\ref{fig:pairMomDistr_vs_time}(a). At $t=0$, 
there is  a
maximum at $k=0$, which, after the quench and a fast transient time that is shorter than $1/J$, moves to 
finite momenta and then approaches $k=q$, where $q$ is given by Eq.~\eqref{eq:qnp}.

\begin{figure}[t]
\includegraphics[width=.96\columnwidth]{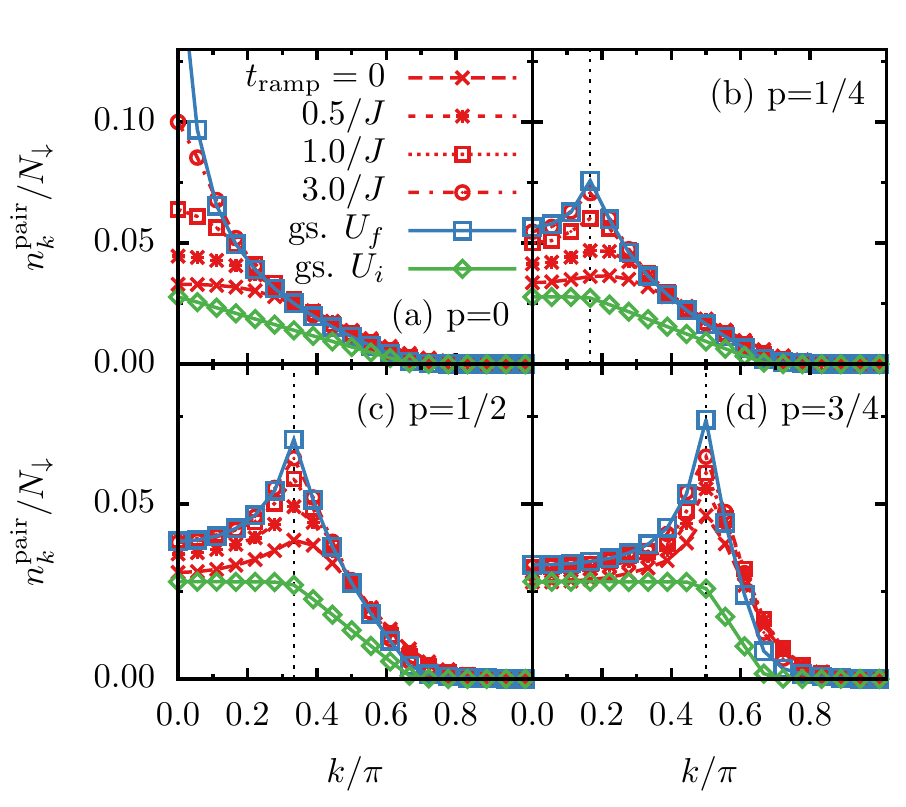}
\caption{(Color online) Time average  of the pair-momentum distribution $n_k^\mathrm{pair}$ between $t=t_\mathrm{ramp}$ and $t_\mathrm{max}=4/J$ after linear ramps
from the noninteracting system ($U_i=0$) to attractive interactions ($U_f=-8J$), for different ramp times $t_\mathrm{ramp}$ and polarizations (symbols with dotted lines):
(a) $p=0$; (b) $p=1/4$; (c) $p=1/2$; (d) $p=3/4$. We also include the data for the ground state of the noninteracting system (diamonds) and the ground state of the attractively interacting system
(squares).  Dotted vertical lines show the maximum position at $k=q$. All results are for $n=2/3$ and $L=36$ (DMRG).
}\label{fig:nkpair_averaged_ramp}
\end{figure}

From its definition, Eq.~\eqref{eq:pmdf}, it follows that the pair MDF is normalized to the double occupancy $\sum_k n_k^{\rm pair} = d$.   Since 
the quench both introduces a redistribution of weight and generates additional pairs, as is evident in
Fig.~\ref{fig:do_avrg_vs_interaction}, the sum over $n_k^\mathrm{pair}$ increases. 
Figures~\ref{fig:pairMomDistr_vs_time}(b) and \ref{fig:pairMomDistr_vs_time}(c) show the individual time evolution of $n^{\rm pair}_{k=q}$ and $n^{\rm pair}_{k=0}$, respectively. 
Both numbers are always below the expectation values in the postquench ground state (dotted lines).
There are  only small finite-size effects, comparing $L=12$ (dashed lines) and $L=36$ (solid lines).
Compared to its initial value at $t=0$, the long-time average of $n_{k=0}^\mathrm{pair}$ does not decrease much (this can be different for other parameter values), while $n_{k=q}^\mathrm{pair}$ increases significantly. 
In any case, the largest absolute increase in $n_{k}^\mathrm{pair}$ typically occurs at $k=q$ (compare also Fig.~\ref{fig:nkpair_averaged_ramp}).

\begin{figure}[t]
\includegraphics[width=.96\columnwidth]{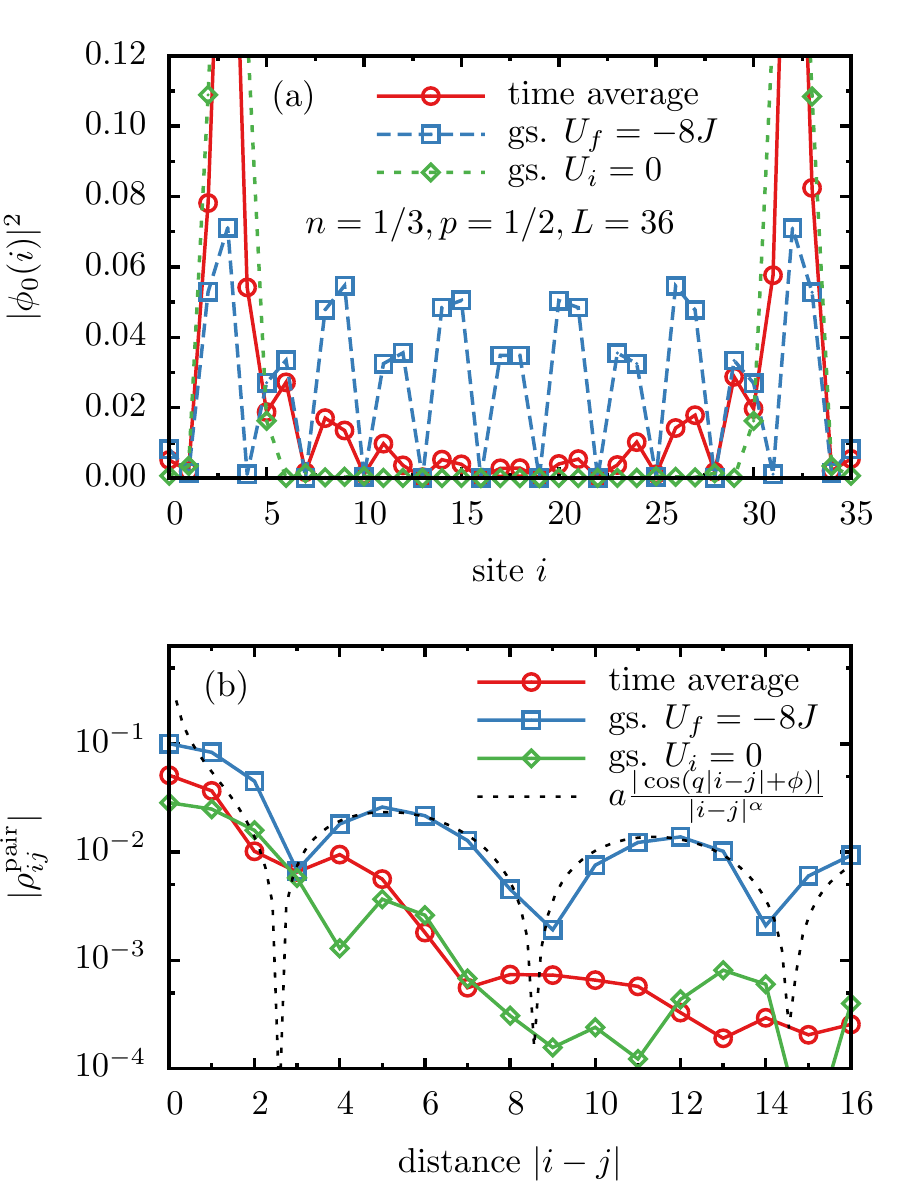}
\caption{(Color online) 
(a) Lowest natural orbital $|\phi_0(i)|^2$  in the  ground state of the noninteracting system ($U_i=0$; diamonds), and in the ground state of the attractively interacting system ($U_f=-8J$; squares), and time average (for $t>3/J$) after the initial quench dynamics (circles). 
(b) Time-averaged spatial decay (circles) of pair-pair correlations $|\rho_{ij}^\mathrm{pair}|$ after a quench from $U_i=0$ to $U_f=-8J$ (circles; time averages  calculated for $t>3/J$), in the  ground state of the noninteracting system ($U=0$; diamonds), and in the ground state of the attractively interacting system ($U=-8J$; squares). Dotted line: Fit  of $f(|i-j|)=a\frac{|\cos(q|i-j|+\phi)|}{|i-j|^{\alpha}}$ to the $U=-8J$ ground state for $|i-j|\ge 3$, with $q$ taken from the position of the maximum in the pair-momentum distribution and fitting parameters $a,\alpha,$ and $\phi$. 
All data are for $n=1/3$, $p=1/2$, and $L=36$ (DMRG).
}\label{fig:corrDecay_natorb_averaged}
\end{figure}

{\it Natural orbitals and decay of correlations.}
In order to further characterize the state after the fast quench dynamics, it is instructive to 
compute the natural orbitals $\phi_\mu(i)$, which are the eigenvectors of the pair-pair correlation function $\rho^{\rm pair}_{ij}$ defined in Eq.~\eqref{eq:rhoij}.
We are, in particular, interested in the spatial structure of the natural orbital $\phi_0(i)$ that belongs to the largest eigenvalue of  $\rho^{\rm pair}_{ij}$.
The time average of this state [i.e., the time average over $|\phi_0(i)|^2$] is plotted  in Fig.~\ref{fig:corrDecay_natorb_averaged}(a),
compared to the same quantity in the postquench and prequench ground state (all for $U_f=-8J$).
As expected we observe a spatial oscillation with $k=q$ that was not present in the initial state (diamonds).
The amplitude in the time-averaged state is smaller than in the postquench ground state (squares).

\begin{figure}[t]
\includegraphics[width=.96\columnwidth]{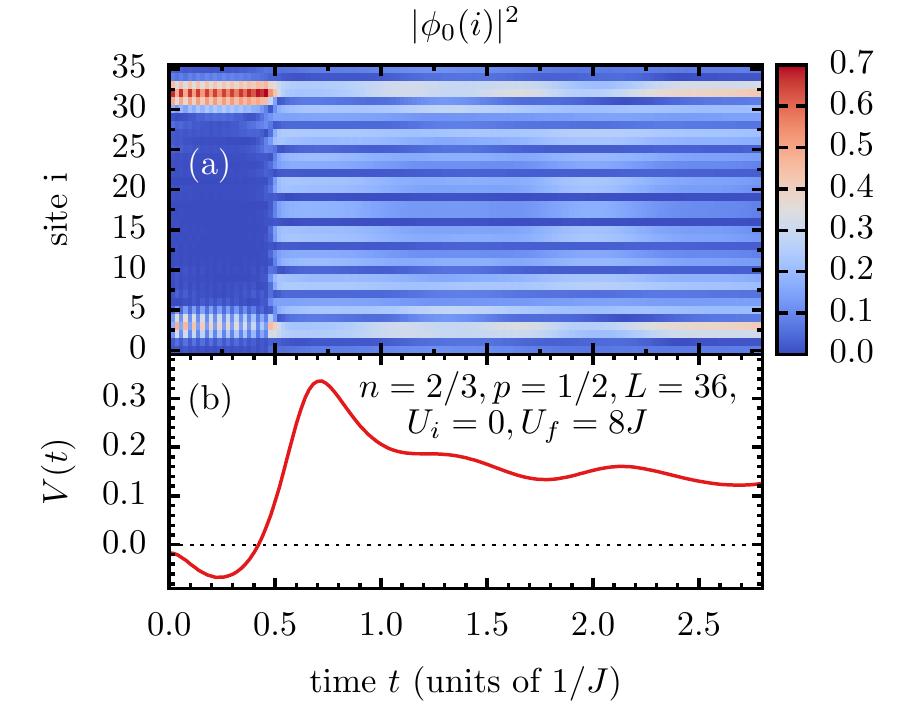}
\caption{(Color online) (a) Time evolution of the lowest natural orbital $|\phi_0(i)|^2$ [for better color contrast, we plot $|\phi_0(i)|$] and (b) time evolution of the FFLO  visibility $V$ [see Eq.~\eqref{eq:visibility}].  The time at which the natural orbital $|\phi_0(i)|^2$ develops an oscillatory pattern roughly coincides with the time at which the visibility goes through 0.
For the prequench system with $U_i=0$ the lowest natural orbital is twofold degenerate, where one of them has a deviation from the constant background
near the left, and the other near the right, boundary. Numerically, we select one of the degenerate orbitals, which explains the spatial anisotropy of $|\phi_0(i)|^2$  for $t\lesssim \tau_\mathrm{FFLO}$. All data for $n=2/3$, $p=1/2$, and $L=36$ (DMRG). 
}\label{fig:natOrb_vis_vs_time2}
\end{figure}

The pair correlation function $\rho^{\rm pair}_{ij}$ itself, shown in Fig.~\ref{fig:corrDecay_natorb_averaged}(b), has a power-law decay in
the postquench ground state superposed with oscillations \cite{yang01,feiguin07}. The same characteristic FFLO oscillations 
emerge in the time-averaged postquench state, yet the decay of the spatial correlations is much faster and roughly exponential.

From these results, we conclude that the time-averaged state indeed has the characteristic features of the FFLO state, albeit not
with quasi-long-range order, which is  expected since the system is at a finite energy above the ground state.
Our next goal is to define and analyze a characteristic time scale for the formation of the FFLO correlations.

{\it Time scale for the formation of FFLO correlations.}
In contrast to the double occupancy, 
$n_k^{\rm pair}(t)$ does not exhibit a single relaxation time, i.e., it is not well approximated by a simple fit function of the type of Eq.~\eqref{eq:fit_d}.
In order to, nevertheless, define a time scale for the formation of FFLO correlations, we find it instructive to
study the time evolution of the natural orbital $\phi_0(i)$.
This quantity is shown in Fig.~\ref{fig:natOrb_vis_vs_time2}(a). Initially, apart from oscillations in $|\phi_0(i)|^2$ induced by the open
boundaries, $\phi_0(i)$ is flat, yet at times $t\sim 0.5 J$, it clearly develops a spatially modulated pattern, indicative of FFLO correlations.

\begin{figure}[t]
\includegraphics[width=.96\columnwidth]{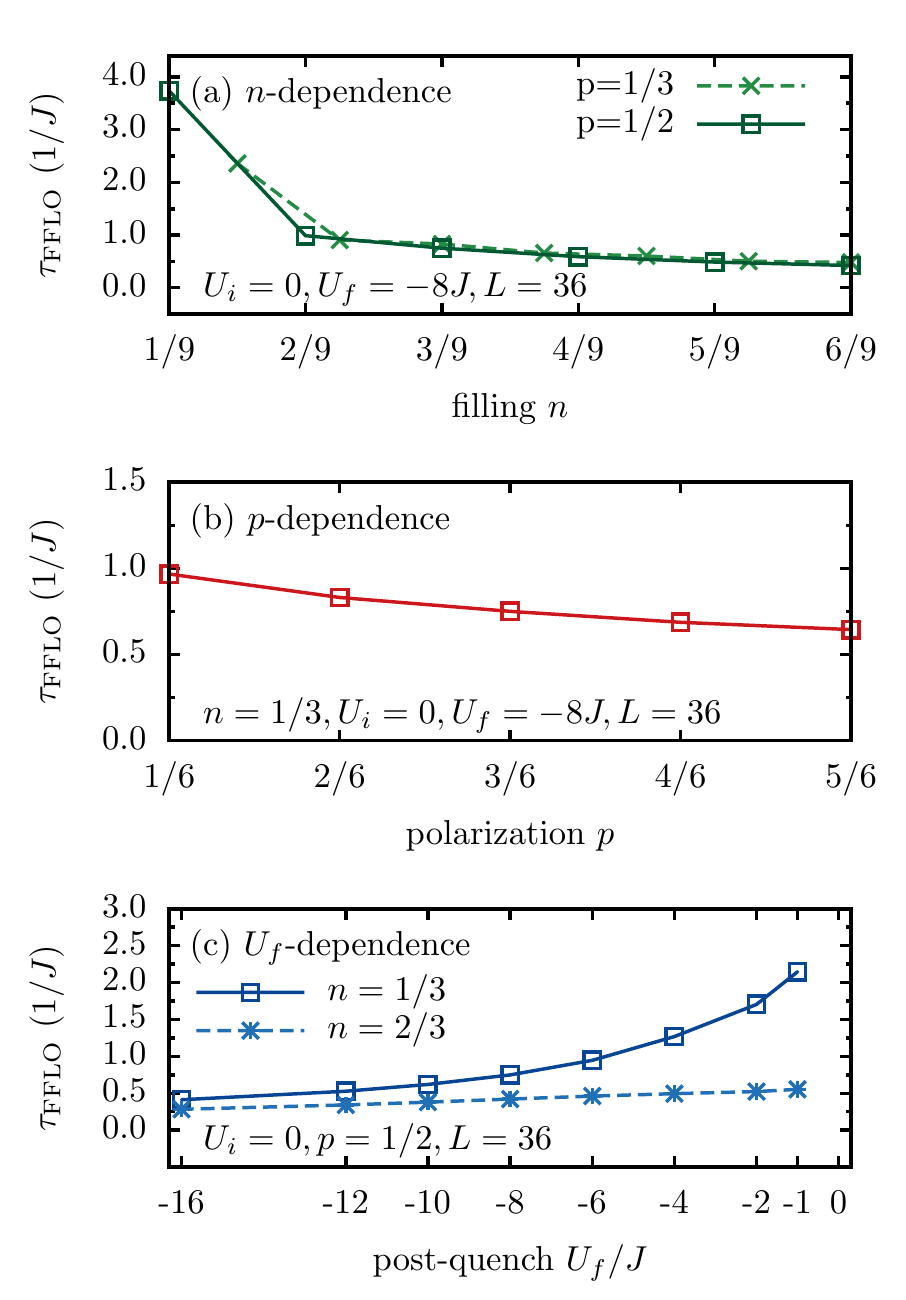}
\caption{(Color online) Time scale $\tau_{\rm FFLO}$ for the formation of FFLO correlations in the quench from $U_i=0$ to $U_f<0$,  versus (a)  filling $n$, (b)  polarization $p$, and (c)  postquench interaction strength $U_f$, all for $L=36$ (DMRG). The time scale is extracted as the first 0 of the visibility $V(t)$.
}\label{fig:vis_buildup}
\end{figure}

The time $\tau_{\rm FFLO}$ at which this modulation appears in $|\phi_0(i)|^2$ is also reflected in the time dependence of the visibility $V(t)$ of the FFLO peak in $n_k^{\rm pair}$
defined in Eq.~\eqref{eq:visibility}, namely, for $t\lesssim \tau_{\rm FFLO}$, $V(t) <0$ while for $t\gtrsim \tau_{\rm FFLO}$, $V(t)>0$, as is evident from Fig.~\ref{fig:natOrb_vis_vs_time2}.
We therefore define the time scale characteristic for the formation of the FFLO correlations as the (first) 0 of the visibility,
\begin{equation}
V(\tau_{\rm FFLO}) =0\,.
\end{equation} 
Note that this could  be measured in experiments, provided the pair MDF is accessible.

Our results for the dependence of $\tau_{\rm FFLO}$ on filling, polarization, and postquench interaction strength are summarized in Figs.~\ref{fig:vis_buildup}(a), \ref{fig:vis_buildup}(b), and \ref{fig:vis_buildup}(c), respectively.
The overall density dependence is similar to that of $\tau_d$ discussed in Sec.~\ref{sec:d_quench1} (compare Fig.~\ref{fig:do_buildup}):
starting from low densities, $\tau_{\rm FFLO}$ decreases. 
It also monotonously decreases with $p$, yet exhibits a 
 weaker overall variation. 

The  processes behind the formation of the FFLO correlations and the formation of the oscillatory structure in the
natural orbital $|\phi_0(i)|^2$ are  the pair formation, redistribution of momentum among pairs, and spatial redistribution of excess  fermions. The time scale for
pair formation is similar to that for the formation of doublons [in excess of those present in the initial state; see Eq.~\eqref{din}], 
which are essentially a measure of the pairs, while the excess fermions have to
rearrange themselves over distances of $l \sim 1/(2q)$. 

\begin{figure}[t]
\includegraphics[width=.96\columnwidth]{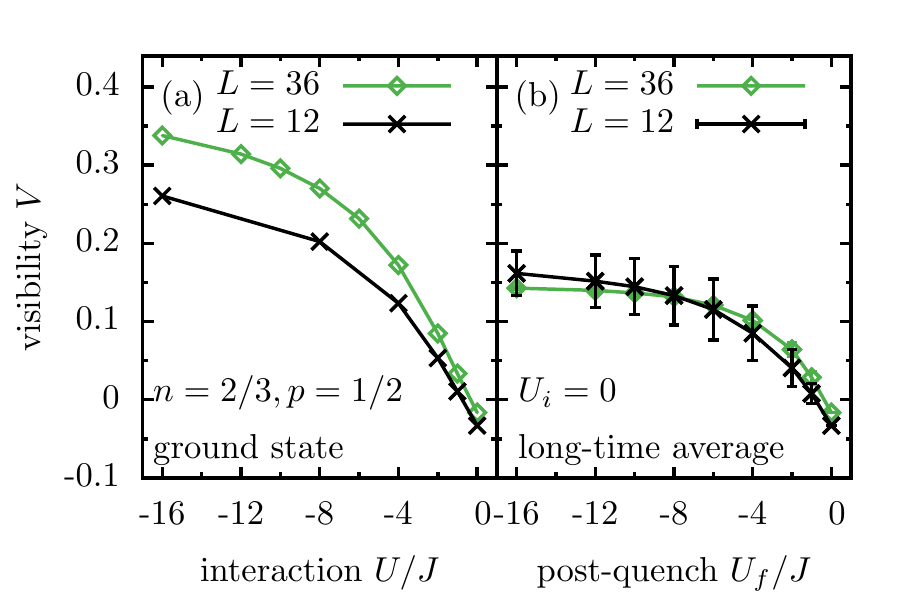}
\caption{(Color online) (a) Visibility $V$ of the FFLO peak in $n_k^{\rm pair}$  in (a) the ground state at interaction strength $U$, and (b) the long-time average (calculated for $t>3/J$) of $V$ versus the postquench interaction strength $U_f$ after  the quench from $U_i=0$, all for $L=36$ (DMRG) and $L=12$ (ED). Error bars indicate the standard deviation from 
the time average (not visible for $L=36$).
}\label{fig:vis_vs_interaction}
\end{figure}

The probability of a minority fermion  finding an excess fermion is
proportional to the density of majority fermions $n_{\uparrow}=N_{\uparrow}/L$. Therefore,  we expect a leading dependence for the formation of additional pairs
given by $\tau_{\rm FFLO} \propto 1/n_{\uparrow} = \frac{1}{n (1+p)}$.
Our numerical results for $\tau_{\rm FFLO}$ extracted from the 0 of the visibility shown in Fig.~\ref{fig:vis_buildup} are in reasonable qualitative agreement with this 
simple argument. 

There is  also an interesting dependence on $U_f$ [see Fig.~\ref{fig:vis_buildup}(c)]: (i) the smaller $|U_f|$, the larger $\tau_{\rm FFLO}$, and (ii) the lower the density, the more strongly $\tau_{\rm FFLO}$ depends on $U_f$.
The first aspect is  intuitive, since for 
$|U_f|\to 0 $,
no FFLO correlations are ever formed, and hence $\tau_{\rm FFLO} \to \infty$.
Note also that for large $|U_f|/J$, $\tau_{\rm FFLO}$ seems to become independent of the density.

{\it Time average of the visibility.}
We finally discuss the long-time average of the visibility as a function of the postquench interaction strength.
The results are shown in Fig.~\ref{fig:vis_vs_interaction}, for the dependence of $V$ both on $U/J$ in the ground state [Fig.~\ref{fig:vis_vs_interaction}(a)] and on the postquench interaction 
strength $U_f/J$ [Fig.~\ref{fig:vis_vs_interaction}(b)]. 
In both cases, $V$ increases monotonically with $U$ and $U_f$, respectively, however, the comparison of different system sizes shows 
only very small variations for the time averages, while in the ground state, $V$ increases with $L$. This is consistent with the fact that in
the ground state we have an actual quasicondensate.

\subsection{Ramps from $U=0$ to $U<0$: Crossover to adiabatic dynamics}
\label{sec:ramps1}

To conclude our discussion of the real-time dynamics starting from a noninteracting gas, we turn to 
slow quenches with a linear ramp from $U_i=0$ to $U_f$ according to Eq.~\eqref{eq:ramp}.
We are, in particular, interested in the crossover to adiabatic behavior, for which we expect to
obtain the ground-state expectation values of observables in the time average. 

Our results for the time average of $n_k^{\rm pair}$ versus quasi-momentum $k$ for ramps and quenches to $U_f=-8J$ 
are presented in Figs.~\ref{fig:nkpair_averaged_ramp}(a), \ref{fig:nkpair_averaged_ramp}(b), \ref{fig:nkpair_averaged_ramp}(c), and \ref{fig:nkpair_averaged_ramp}(d) for polarizations $p=0$, $1/4$, $1/2$, and $3/4$, respectively.
These plots also include $n_k^{\rm pair}$ calculated in the ground state at $U_i=0$ and $U_f=-8J$ and the results for
instantaneous quenches. For the quenches, the time averages at $p>0$ result in maxima at $k=q$ but the distribution
is mostly reduced in height compared to the postquench ground state (squares).
  
Moreover, we see that,  for both quenches and ramps, for large values of the quasi-momentum $k$, the time-averaged  pair momentum distribution is already 
very close to its ground-state value for $U=U_f$ and therefore does not depend much on the sweep rate. For small quasi-momenta,
the averaged  pair momentum distribution increases by decreasing the sweep rate. Nevertheless, the comparably short ramp time of $t_{\rm ramp}\sim 3/J$ is enough to obtain time averages that are very close to the ground-state correlations also for small quasi-momenta.

It is important to emphasize that  the time-averaged  pair momentum distribution, for most parameter values, exhibits the typical FFLO peak at 
$k=q$, which is expected in the ground state for attractive interactions, $U=U_f<0$.

\subsection{Summary: Quenches and linear ramps from $U_i=0$ to $U_f<0$}
As a main result of our analysis of quantum quenches from $U_i=0$ to attractive interactions, we showed
that the time average of the double occupancy has a maximum at $|U_f| \sim 4J$, while the value at small 
$|U_f|/J$ is close to the postquench ground-state values. At large $|U_f|/J$, the initial double occupancy simply
gets frozen in on the accessible time scales.
Interestingly, we observe that
the approach to the stationary regime is exponential in time. 
This relaxation process is superimposed with coherent oscillations with  $\omega \sim |U_f|$ for large $|U_f|/J$.

Even though the quantum quench puts the system at high energies, the pair MDF still 
develops a maximum at the characteristic FFLO momentum $q$. We analyzed the
eigenstates of the pair density matrix $\rho_{ij}^{\rm pair}$ and showed that the emergence of a maximum in $n_k^{\rm pair}$ is accompanied 
by the eigenstate of $\rho_{ij}^{\rm pair}$ belonging to the largest eigenvalue developing a spatially oscillatory structure.
This  coincides with the point in time 
at which the visibility of the FFLO peak in $n_k^{\rm pair}$ becomes positive, meaning that the maximum is at $k=q$. 
We therefore extracted a time scale for the formation of oscillatory pair correlations from the first 0 of the visibility.
In ramps from $U_i=0$ to $U_f<0$, we observe that ramp times of $t_{\rm ramp} \sim 3/J$  are enough 
to be adiabatic in the sense that the time average of $n_k^{\rm pair}$ after the ramp cannot be discerned from the 
postquench ground-state form of $n_k^{\rm pair}$. Overall, both quenches and ramps produce the largest visibilities 
of the FFLO maximum in $n_k^{\rm pair} $at large polarizations (compare Fig.~\ref{fig:nkpair_averaged_ramp}).

It is tempting to relate the time averages to a temperature by comparison with thermodynamic ensembles, as is often done in 
studies of thermalization in interaction quantum quenches (see \cite{sorg14} and references therein). 
The Fermi-Hubbard model being integrable, we expect that  an infinitely large number of Lagrange parameters (order of $L$ many)
is necessary to describe the time averages with a generalized Gibbs ensemble along the lines of \cite{rigol07}.
While it is not the purpose of this study to clarify the validity of the generalized Gibbs ensemble for this particular quench, 
we note that we verified that the time averages of $n_k^{\rm pair}$ cannot be described by the canonical ensemble, as expected for an integrable model. 

Various  extensions of our analysis such as the
formation of a polaron (i.e., $N_\downarrow=1$), the dynamics in the low-density limit, and the dynamics starting from
product states constitute interesting problems for future research. 
Some of these questions may permit an analytical
solution and product states are commonly studied in quantum gas experiments (namely, the dynamics starting from product states in real space, see, e.g., \cite{trotzky12,ronzheimer13,cheneau12,perthot14}).

For initial product states, the dynamics can be independent of the sign of $U_f$ \cite{schneider12,iyer13}, provided the observable is
invariant under both a $\pi$ boost and time reversal. 
The latter is not true for the pair MDF, and hence, the quench dynamics of $n_k^{\rm pair}$ can depend on the sign of $U_f$
for certain initial product states. For instance, this happens 
if the distributions of spin-up and spin-down fermions in the lattice break (spatial) inversion symmetry.


\section{Imprinting FFLO correlations onto repulsively bound pairs}
\label{sec:imprint}

\begin{figure}[!t]
\includegraphics[width=.96\columnwidth]{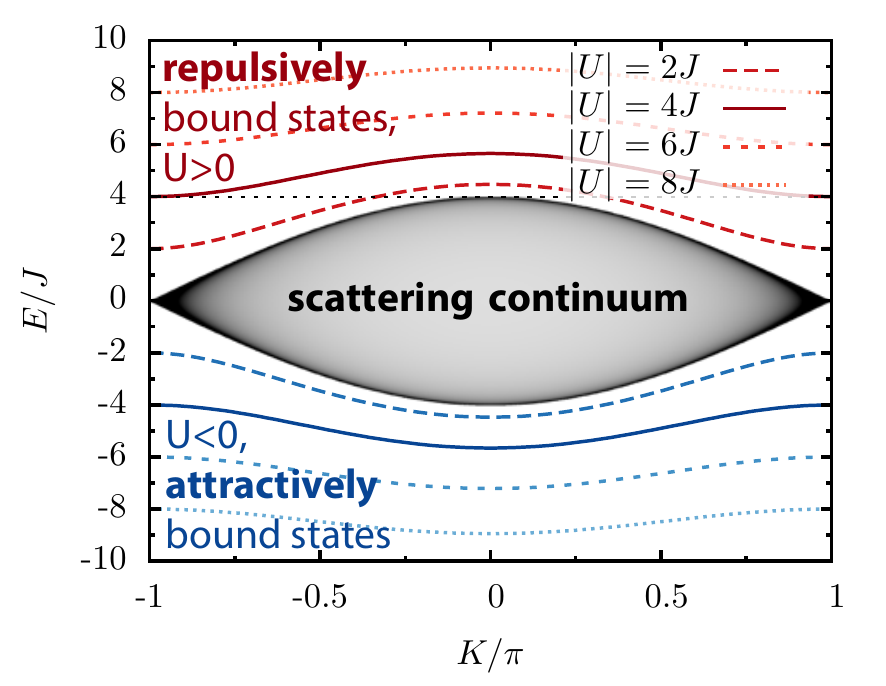}
\caption{(Color online) Spectra of the scattering continuum and the bound states for different on-site interaction strengths $U$ from a two-particle  calculation ($N_\uparrow=N_{\downarrow}=1$) for pairs with total momentum $K$ in the  Fermi-Hubbard model. The  shading of the scattering continuum is proportional to the density of states (DOS); darker gray indicates a higher DOS (note that the DOS diverges at the edges of the continuum region). The  bandwidth of the scattering continuum is $8J$. Bound states are discrete and symmetric for $U\rightarrow -U$. There is no
 energy overlap of repulsively bound states with the continuum for $U>4J$ (dotted black line).
}
\label{fig:sketch_pairs}
\end{figure}

By carrying out a quench from negative to positive values of $U$, we aim at imprinting FFLO correlations onto
repulsively bound pairs present on the $U>0$ side. In one dimension, 
repulsively and attractively bound pairs exist for all values of $U>0$ and $U<0$, respectively.
Figure~\ref{fig:sketch_pairs} shows a sketch of the spectrum of repulsively and attractively bound pairs.
The former were observed in a seminal experiment with bosons in optical lattices \cite{winkler06}.
We expect that, if $U_f \gg 4J$, the repulsively bound pairs will be very long-lived, implying that 
FFLO correlations will be preserved for a long time. This is related to the observation that in Hubbard models,
the double occupancy is approximately conserved for  values of $U/J$ larger than the band-width \cite{rosch08},
which is at the heart of many intriguing transient and meta-stable phenomena in quantum quenches such as quantum distillation \cite{hm09,bolech12,muth12,weiss} 
and the exponentially long thermalization times for quenches into the large-$U/J$ regime that involve finite densities of doublons \cite{eckstein11,alhassanieh08,kollath07}.

We first discuss quenches from $U_i<0$ to $U_f>0$ (Sec.~\ref{sec:quench2}) and then turn our attention to linear ramps (Sec.~\ref{sec:ramp2}).

\begin{figure}[t]
\includegraphics[width=.96\columnwidth]{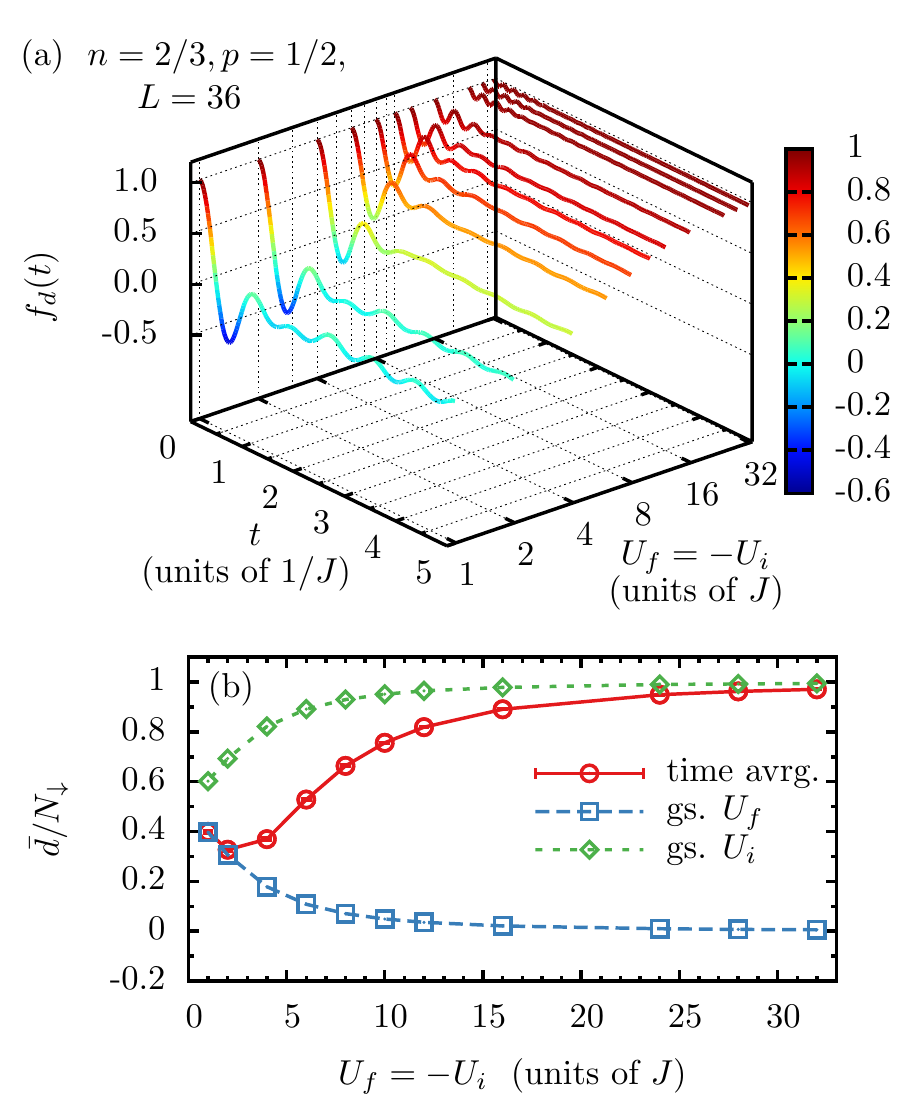}
\caption{(Color online) (a) Relative fraction of excess double occupancy $f_d(t)$ [see Eq.~\eqref{eq:fo}] after the interaction quench from $U_i<0$ to repulsive $U_f=|U_i|>0$, normalized to the difference between post- and prequench ground-state values. 
(b) Time-averaged double occupancy $\bar d$ (circles; for $t>4/J$), postquench ground-state (squares), and initial state (diamonds) versus postquench interaction strength $U_f/J$. All data are for $n=2/3$, $p=1/2$, and $L=36$ (DMRG).}
\label{fig:waterfallplot_interaction}
\end{figure}

\subsection{Interaction quench from $U<0$ to $U>0$}
\label{sec:quench2}
\subsubsection{Double occupancy}
On the attractive side, because of pair formation, a large double occupancy is favored, whereas $U/J>0$ suppresses $d$ in the ground state.
The fraction of quench-induced doublons that survive in the quench from $U_i<0$ to $U_f=|U_i|>0$ is therefore a good measure for the stability
of repulsively bound pairs.
Figure~\ref{fig:waterfallplot_interaction}(a) shows DMRG results for the time evolution of the relative fraction of doublons,
which we define as ($o$ representing any observable)
\begin{equation}
f_o(t)= \frac{o(t) -o_{\rm gs}}{o_0-o_{\rm gs}}\,,
\label{eq:fo}
\end{equation}
where $o_{\rm gs }$ is the expectation value in the postquench ground state and $o_0=o(t=0)$ is the value in the prequench ground state.
Considering $o=d$, a value of $f_d=1$ means that all doublons survive the quench.
The figure unveils that for all  values of $U_i$, after a short transient, the double occupancy settles into a constant, superposed with oscillations.
Moreover, the larger $|U_i|/J$, the more doublons survive, and $f_d\to 1$ for $|U_i|/J \gg 4$.
Note that typical quantum gas experiments with interacting atoms in optical lattices reach time scales of the order of $tJ \lesssim 20$ \cite{trotzky12}, hence
our results apply to the experimentally accessible time window.

Figure~\ref{fig:waterfallplot_interaction}(b) shows the time average $\bar d$ of the double occupancy after the quench, compared
to the initial value $d(t=0)$ and the postquench ground-state value $d_{\rm gs}$. 
$\bar d$ exceeds $d_{\rm gs}(U_f)$ and  approaches $d_{\rm gs}(U_i)$  for $|U_i|\gg 4J$, at least on the accessible time scales. 
The minimum of $\bar d$ at small $|U_f|/J$ has an origin similar to that of the the maximum of $\bar d$ in the quenches from $U_i=0$ to $U_f<0$
(compare Fig.~\ref{fig:do_avrg_vs_interaction}): for low excess energies, $\bar d$ is only slightly above the postquench ground-state values, but ultimately
approaches the value set by the initial conditions for large $|U_f|/J$.

\begin{figure}[t]
\includegraphics[width=.96\columnwidth]{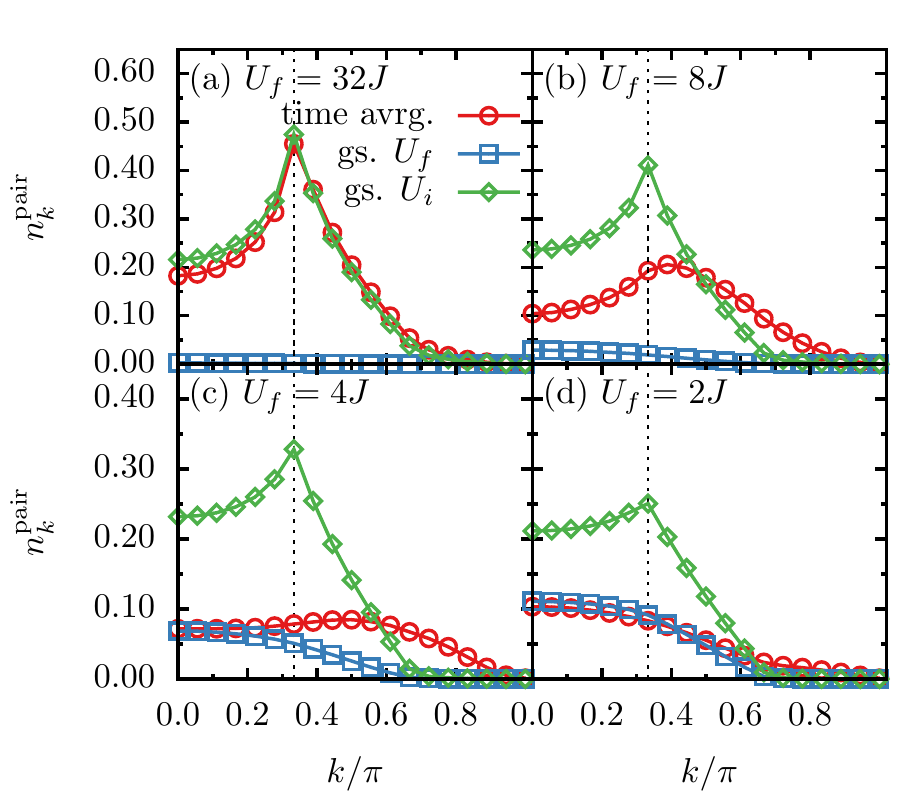}
\caption{(Color online)  Time average of the pair momentum distribution function $n_k^\mathrm{pair}$,  calculated for $t>3/J$, for  the quench from attractive interactions $U_i<0$ to repulsive interactions $U_f=-U_i>0$, for different $U_f$---(a) $U_f/J=32$, (b) $U_f/J=8$, (c) $U_f/J=4$, and (d) $U_f/J=2$ (circles); for  the ground state of the attractive system (diamonds);
and for the ground state of the repulsive system at $U_f$ (squares). All data  for $n=2/3$, $p=1/2$, and $L=36$ (DMRG).
}\label{fig:nkpair_averaged_inter_repulsive}
\end{figure}
 
\subsubsection{Pair momentum distribution function}
Our results for the pair MDF in the quench from $U_i<0$ to $U_f=-U_i$ are presented in Fig.~\ref{fig:nkpair_averaged_inter_repulsive},
which shows the time averages of $n_k^{\rm pair}$ versus $k$ for the four values $U_i/J=-32$, $-8$, $-4$, and $-2$ at polarization $p=1/2$ (circles). The figure also includes the respective $n_k^{\rm pair}$
curves in the pre- and postquench ground states (diamonds and squares, respectively). 
As expected, for large $|U_i|/J \gg 4$
the initial $n_k^{\rm pair}$ is perfectly preserved after the quench (at least for $tJ \leq 5$).
Upon decreasing $|U_i|/J$, the overall height of $n_k^{\rm pair}$ decreases and the maximum in $n_k^{\rm pair}$ shifts towards larger values of $k>q$. This is related  to the fact that for large quasi-momenta the time-averaged pair momentum distribution is always larger than its prequench counterpart.

As the interaction strength becomes of the order of the bandwidth, $|U_i|/J=4$,
we see that the time-averaged  $n_k^{\rm pair}$ approaches the postquench ground-state distribution also for small
momenta. By further decreasing the interaction at $|U_i|/J=2$,  the pair momentum distribution becomes already practically identical to the postquench ground-state curve for all quasi-momenta.

\subsection{Slow quenches: Linear ramps from $U_i<0$ to $|U_i|>0$}
\label{sec:ramp2}

In any experiment, a quench is a fast parameter change which nonetheless takes place over a  finite time window, such that one has to worry about 
quench-type  behavior versus adiabatic changes.
To account for this, we investigate slow quenches with linear ramps [see Eq.~\eqref{eq:ramp}] from $U_i<0$ to $U_f=-U_i$.

\begin{figure}[t]
\includegraphics[width=.96\columnwidth]{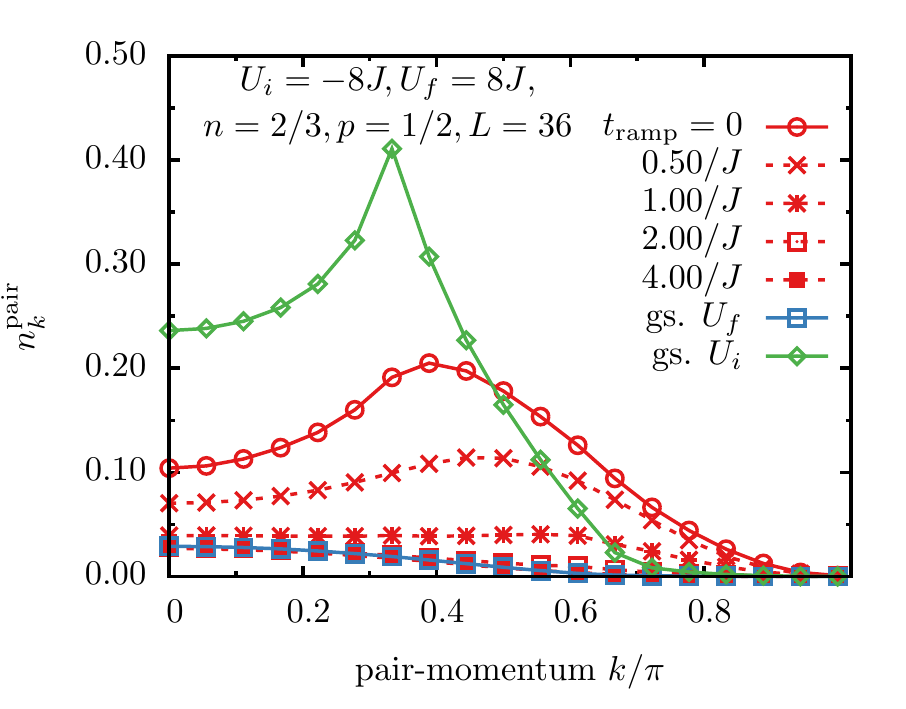}
\caption{(Color online) Time average of the pair-momentum distribution $n_k^\mathrm{pair}$  (calculated for $t>4/J$)  after ramps from attractive interactions ($U_i<0$) to repulsive interactions $U_f=|U_i|>0$, for different ramp times $t_\mathrm{ramp}$ (symbols with dashed lines). We also include results for the instantaneous quench (circles), the ground state of the attractive system at $U_i$ (diamonds),
and the ground state of the repulsive system (squares) for comparison. All results are for $n=2/3$, $p=1/2$, and $L=36$ (DMRG).
}\label{fig:nkpair_averaged_ramp_repulsive}
\end{figure}

An example for the time average of the pair MDF  (calculated after the end of the ramp) is shown in Fig.~\ref{fig:nkpair_averaged_ramp_repulsive},
for various ramp times compared to the (instantaneous) quench and the  expectation values in the pre- and postquench ground state (for $U_i=-8J$).
In this example, the quench already leads to a significant reduction of $n_k^{\rm pair}$ (circles) compared to the initial state (diamonds), and 
a finite ramp time further suppresses the maximum in $n_k^{\rm pair}$ at finite momentum, plus shifts the position of the maximum to larger values. 
Already at $t_{\rm ramp}=2/J$, the time average coincides with the postquench ground-state expectation values (compare the different curves with squares).
It is interesting to note that as the ramp time increases, the resulting pair momentum distribution approaches the postquench
 ground-state distribution  for small momenta  first, in contrast to Fig.~\ref{fig:nkpair_averaged_ramp}, where the quench is from
 the noninteracting to the attractive regime.

The tendency of these quenches and ramps to dominantly populate repulsively bound pairs with large momenta 
can be understood from the fact that the minimum of the dispersion of the (anti-)bound state 
is at $k=\pi$ (compare Fig.~\ref{fig:sketch_pairs}).
Note, though, that $k=\pi$ never gets populated; in fact, $n_{k=\pi}^{\rm pair}$ also vanishes in the prequench state.
In other words, the so-called $\eta$ condensate \cite{yang,rosch08,kantian10}, an exact eigenstate of the Fermi-Hubbard model
with a macroscopic occupation of doublons (or pairs) at center-of-mass momentum $k=\pi$, has no overlap with our
initial state. 

\begin{figure}[t]
\includegraphics[width=.96\columnwidth]{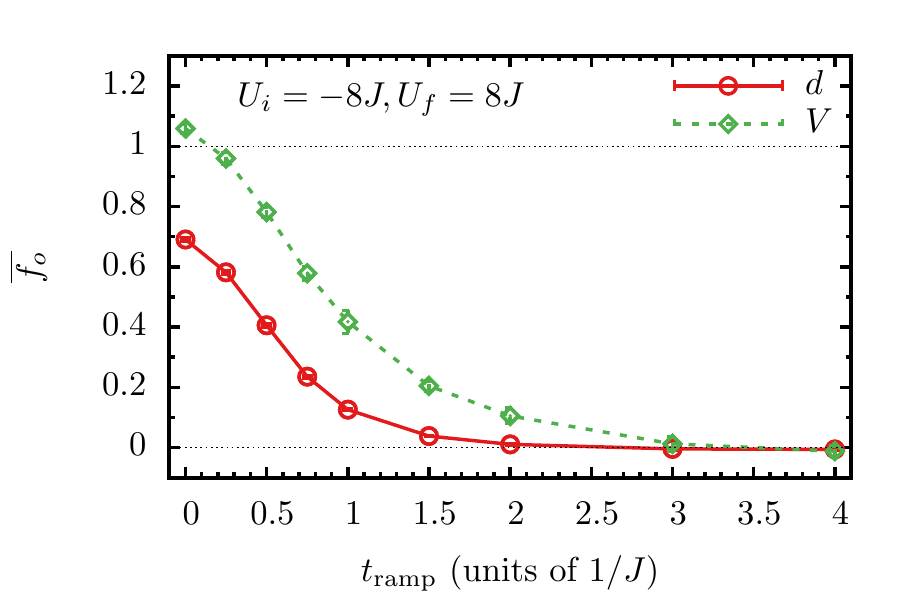}
\caption{(Color online) Time average of $f_o$ [see Eq.~\eqref{eq:fo}] for visibility (diamonds) 
and double occupancy (circles), calculated for $t>4/J$,
versus ramp time $t_\mathrm{ramp}$ after the quench and ramps
from attractive $U_i<0$ to repulsive interactions $U_f = -U_i >0$. All data for $n=2/3$, $p=1/2$, and $L=36$ (DMRG). }\label{fig:ramp_relative}
\end{figure}

Our results for $\bar{f}_d$ and $\bar{f}_V$ [see Eq.~\eqref{eq:fo}] are displayed in Fig.~\ref{fig:ramp_relative}, as a function of the ramp time for $U_i=-8J$, $n=2/3$, $p=1/2$.
Both $\bar{f}_d$ and $\bar{f}_V$ decrease monotonously with $t_{\rm ramp}$ and a value of $t_{\rm ramp} \sim 2/J$ is enough to suppress any 
excess pairs in the time average after the ramp, while the relative visibility $f_V$ remains non-zero for $t_{\rm ramp} \lesssim 3/J$ for these parameters.  
Note, though, that the maximum of $n_k^{\rm pair}$ is not necessarily at $k=q$ in the postquench state [see Fig.~\ref{fig:nkpair_averaged_ramp_repulsive}].

\section{Summary and Conclusion}
\label{sec:sum} 
In this work we have presented an extensive numerical analysis of quantum quenches and linear ramps
of the interaction strength in the 1D Fermi-Hubbard model. We are particularly interested in
quenches that involve the FFLO state: we have considered quenches from the noninteracting case to negative
values of $U$ at finite spin imbalances as well as quenches from negative to positive $U$.

For the first example, we present results for the double occupancy and the pair MDF. We highlight four key observations: (i) The time average of the postquench
double occupancy takes a maximum at $|U_f|/J \sim 4$; (ii) there are parameter regimes where the relaxation towards the steady state is exponential;  (iii) the emergence of FFLO correlations in the postquench state leaves clear fingerprints in the pair MDF; and (iv) we further demonstrate that linear ramps with ramp times of $3/J$ are sufficient to be essentially adiabatic. These aspects should, in principle, be accessible in experiments. The current obstacle is that an experimental measurement of the pair momentum distribution has not yet been accomplished for 1D quantum gases. The measurement of the double occupancy is a well-established technique.

In the second part of the paper, where we consider quenches from negative to positive values of the interaction
strength, we have demonstrated that FFLO correlations can be imprinted onto repulsively bound pairs if the
postquench interaction strength is high. This generates a metastable state due to the exponentially long
life-time of doublons generated in such quenches \cite{rosch08}.
In linear ramps, one quickly approaches the adiabatic case with ramp times as short as $t_{\rm ramp}\sim 2/J$
and FFLO signatures are rapidly lost.

There are several interesting extensions of our work. For instance, for quenches of a spin-imbalanced gas 
from $U=0$ to $U<0$ in an inhomogeneous system such as realized in a harmonic trap, there need to be 
macroscopic spin currents in the dynamical emergence  of  the  shell structures 
characteristic of a 1D spin-imbalanced Fermi gas with attractive contact interactions  \cite{orso07,hu07,feiguin07,casula08,hm10a}.
In this system, the core of the system is partially polarized in one dimension and in the wings, it is either fully paired or fully
polarized, depending on the global polarization. The small polarization regime should be the most interesting
one. Such an experiment could potentially give access to the transport coefficients of such a system
(see \cite{tezuka10} for work in that direction).

To further elucidate the relaxation dynamics of the Fermi-Hubbard model, the dynamics starting from product states in real space
(see, e.g., \cite{trotzky12}, \cite{perthot14},  and \cite{vidmar13} for first numerical results) and the time-dependent formation of a polaron constitute 
potentially very instructive limiting cases for future studies. 

For the question of thermalization in the Fermi-Hubbard model, where in general one would expect
thermalization with respect to the generalized Gibbs ensemble because of the integrability of the model \cite{rigol07},
at present the complicated structure of the integrals of motion presents an obstacle for testing
this conjecture. The effect of integrability breaking can, in experiments, be studied either by using
two-leg ladders \cite{feiguin09}, realized with superlattices \cite{foelling07}, in fermionic superfluids with resonant interactions \cite{baur10,hm10}, or with mass-imbalanced systems, realizable with
two-component gases of different fermionic species (see the discussion in \cite{batrouni09,wang09,orso10,roux11,dalmonte12}) or using spin-dependent optical lattices \cite{mandel}.
A recent study \cite{moore} discussed quasi-many-body localization in such a system.

{\it Acknowledgment.} We thank F. Essler,  J.-N. Fuchs, L. Radzihovsky, and W. Zwerger for helpful discussions and we thank S. Kehrein for comments on an early version of the manuscript. We further thank C. A. B\"usser for his contributions at early stages of the project.
F. H.-M. was supported by the Deutsche Forschungsgemeinschaft through
DFG Research Unit FOR 801.


\bibliographystyle{apsrev}
\bibliography{references}

\begin{thebibliography}{100}
\expandafter\ifx\csname natexlab\endcsname\relax\def\natexlab#1{#1}\fi
\expandafter\ifx\csname bibnamefont\endcsname\relax
  \def\bibnamefont#1{#1}\fi
\expandafter\ifx\csname bibfnamefont\endcsname\relax
  \def\bibfnamefont#1{#1}\fi
\expandafter\ifx\csname citenamefont\endcsname\relax
  \def\citenamefont#1{#1}\fi
\expandafter\ifx\csname url\endcsname\relax
  \def\url#1{\texttt{#1}}\fi
\expandafter\ifx\csname urlprefix\endcsname\relax\def\urlprefix{URL }\fi
\providecommand{\bibinfo}[2]{#2}
\providecommand{\eprint}[2][]{\url{#2}}

\bibitem[{\citenamefont{Sarma}(1963)}]{sarma63}
\bibinfo{author}{\bibfnamefont{G.}~\bibnamefont{Sarma}},
  \bibinfo{journal}{Phys. Chem. Solids} \textbf{\bibinfo{volume}{24}},
  \bibinfo{pages}{1029} (\bibinfo{year}{1963}).

\bibitem[{\citenamefont{Fulde and {Ferrell}}(1964)}]{fulde64}
\bibinfo{author}{\bibfnamefont{P.}~\bibnamefont{Fulde}} \bibnamefont{and}
  \bibinfo{author}{\bibfnamefont{A.}~\bibnamefont{{Ferrell}}},
  \bibinfo{journal}{Phys. Rev.} \textbf{\bibinfo{volume}{135}},
  \bibinfo{pages}{A550} (\bibinfo{year}{1964}).

\bibitem[{\citenamefont{Larkin and Ovchinnikov}(1964)}]{larkin64}
\bibinfo{author}{\bibfnamefont{A.}~\bibnamefont{Larkin}} \bibnamefont{and}
  \bibinfo{author}{\bibfnamefont{Y.}~\bibnamefont{Ovchinnikov}},
  \bibinfo{journal}{Zh. Eksp. Teor. Fiz} \textbf{\bibinfo{volume}{47}},
  \bibinfo{pages}{1136} (\bibinfo{year}{1964}).

\bibitem[{\citenamefont{Sheehy and Radzihovsky}(2007)}]{sheehy07}
\bibinfo{author}{\bibfnamefont{D.~E.} \bibnamefont{Sheehy}} \bibnamefont{and}
  \bibinfo{author}{\bibfnamefont{L.}~\bibnamefont{Radzihovsky}},
  \bibinfo{journal}{Ann. Phys.} \textbf{\bibinfo{volume}{322}},
  \bibinfo{pages}{1790} (\bibinfo{year}{2007}).

\bibitem[{\citenamefont{Chevy and Mora}(2010)}]{chevy10}
\bibinfo{author}{\bibfnamefont{F.}~\bibnamefont{Chevy}} \bibnamefont{and}
  \bibinfo{author}{\bibfnamefont{C.}~\bibnamefont{Mora}},
  \bibinfo{journal}{Rep. Prog. Phys} \textbf{\bibinfo{volume}{73}},
  \bibinfo{pages}{112401} (\bibinfo{year}{2010}).

\bibitem[{\citenamefont{Gubbels and Stoof}(2013)}]{gubbels13}
\bibinfo{author}{\bibfnamefont{K.}~\bibnamefont{Gubbels}} \bibnamefont{and}
  \bibinfo{author}{\bibfnamefont{H.}~\bibnamefont{Stoof}},
  \bibinfo{journal}{Phys. Rep.} \textbf{\bibinfo{volume}{525}},
  \bibinfo{pages}{255} (\bibinfo{year}{2013}).

\bibitem[{\citenamefont{Radovan et~al.}(2003)\citenamefont{Radovan, Fortune,
  Murphy, Hannahs, Palm, Tozer, and Hall}}]{rado03}
\bibinfo{author}{\bibfnamefont{H.~A.} \bibnamefont{Radovan}},
  \bibinfo{author}{\bibfnamefont{N.~A.} \bibnamefont{Fortune}},
  \bibinfo{author}{\bibfnamefont{T.~P.} \bibnamefont{Murphy}},
  \bibinfo{author}{\bibfnamefont{S.~T.} \bibnamefont{Hannahs}},
  \bibinfo{author}{\bibfnamefont{E.~C.} \bibnamefont{Palm}},
  \bibinfo{author}{\bibfnamefont{S.~W.} \bibnamefont{Tozer}}, \bibnamefont{and}
  \bibinfo{author}{\bibfnamefont{D.}~\bibnamefont{Hall}},
  \bibinfo{journal}{Nature} \textbf{\bibinfo{volume}{425}}, \bibinfo{pages}{51}
  (\bibinfo{year}{2003}).

\bibitem[{\citenamefont{Uji et~al.}(2006)\citenamefont{Uji, Terashima,
  Nishimura, Takahide, Konoike, Enomoto, Cui, Kobayashi, Kobayashi, Tanaka
  et~al.}}]{uji06}
\bibinfo{author}{\bibfnamefont{S.}~\bibnamefont{Uji}},
  \bibinfo{author}{\bibfnamefont{T.}~\bibnamefont{Terashima}},
  \bibinfo{author}{\bibfnamefont{M.}~\bibnamefont{Nishimura}},
  \bibinfo{author}{\bibfnamefont{Y.}~\bibnamefont{Takahide}},
  \bibinfo{author}{\bibfnamefont{T.}~\bibnamefont{Konoike}},
  \bibinfo{author}{\bibfnamefont{K.}~\bibnamefont{Enomoto}},
  \bibinfo{author}{\bibfnamefont{H.}~\bibnamefont{Cui}},
  \bibinfo{author}{\bibfnamefont{H.}~\bibnamefont{Kobayashi}},
  \bibinfo{author}{\bibfnamefont{A.}~\bibnamefont{Kobayashi}},
  \bibinfo{author}{\bibfnamefont{H.}~\bibnamefont{Tanaka}},
  \bibnamefont{et~al.}, \bibinfo{journal}{Phys. Rev. Lett.}
  \textbf{\bibinfo{volume}{97}}, \bibinfo{pages}{157001}
  (\bibinfo{year}{2006}).

\bibitem[{\citenamefont{Casalbuoni and Nardulli}(2004)}]{casal04}
\bibinfo{author}{\bibfnamefont{R.}~\bibnamefont{Casalbuoni}} \bibnamefont{and}
  \bibinfo{author}{\bibfnamefont{G.}~\bibnamefont{Nardulli}},
  \bibinfo{journal}{Rev. Mod. Phys.} \textbf{\bibinfo{volume}{76}},
  \bibinfo{pages}{263} (\bibinfo{year}{2004}).

\bibitem[{\citenamefont{Feiguin et~al.}(2012)\citenamefont{Feiguin,
  Heidrich-Meisner, Orso, and Zwerger}}]{feiguin11}
\bibinfo{author}{\bibfnamefont{A.~E.} \bibnamefont{Feiguin}},
  \bibinfo{author}{\bibfnamefont{F.}~\bibnamefont{Heidrich-Meisner}},
  \bibinfo{author}{\bibfnamefont{G.}~\bibnamefont{Orso}}, \bibnamefont{and}
  \bibinfo{author}{\bibfnamefont{W.}~\bibnamefont{Zwerger}},
  \bibinfo{journal}{Lect. Notes Phys.} \textbf{\bibinfo{volume}{836}},
  \bibinfo{pages}{503} (\bibinfo{year}{2012}).

\bibitem[{\citenamefont{Guan et~al.}(2013)\citenamefont{Guan, Batchelor, and
  Lee}}]{guan13}
\bibinfo{author}{\bibfnamefont{X.-W.} \bibnamefont{Guan}},
  \bibinfo{author}{\bibfnamefont{M.~T.} \bibnamefont{Batchelor}},
  \bibnamefont{and} \bibinfo{author}{\bibfnamefont{C.}~\bibnamefont{Lee}},
  \bibinfo{journal}{Rev. Mod. Phys.} \textbf{\bibinfo{volume}{85}},
  \bibinfo{pages}{1633} (\bibinfo{year}{2013}).

\bibitem[{\citenamefont{Liao et~al.}(2010)\citenamefont{Liao, Rittner,
  Paprotta, Li, Partridge, Hulet, Baur, and Mueller}}]{liao10}
\bibinfo{author}{\bibfnamefont{Y.-A.} \bibnamefont{Liao}},
  \bibinfo{author}{\bibfnamefont{A.~S.~C.} \bibnamefont{Rittner}},
  \bibinfo{author}{\bibfnamefont{T.}~\bibnamefont{Paprotta}},
  \bibinfo{author}{\bibfnamefont{W.}~\bibnamefont{Li}},
  \bibinfo{author}{\bibfnamefont{G.~B.} \bibnamefont{Partridge}},
  \bibinfo{author}{\bibfnamefont{R.~G.} \bibnamefont{Hulet}},
  \bibinfo{author}{\bibfnamefont{S.~K.} \bibnamefont{Baur}}, \bibnamefont{and}
  \bibinfo{author}{\bibfnamefont{E.~J.} \bibnamefont{Mueller}},
  \bibinfo{journal}{Nature} \textbf{\bibinfo{volume}{467}},
  \bibinfo{pages}{567} (\bibinfo{year}{2010}).

\bibitem[{\citenamefont{Orso}(2007)}]{orso07}
\bibinfo{author}{\bibfnamefont{G.}~\bibnamefont{Orso}}, \bibinfo{journal}{Phys.
  Rev. Lett.} \textbf{\bibinfo{volume}{98}}, \bibinfo{pages}{070402}
  (\bibinfo{year}{2007}).

\bibitem[{\citenamefont{Hu et~al.}(2007)\citenamefont{Hu, Liu, and
  Drummond}}]{hu07}
\bibinfo{author}{\bibfnamefont{H.}~\bibnamefont{Hu}},
  \bibinfo{author}{\bibfnamefont{X.-J.} \bibnamefont{Liu}}, \bibnamefont{and}
  \bibinfo{author}{\bibfnamefont{P.~D.} \bibnamefont{Drummond}},
  \bibinfo{journal}{Phys. Rev. Lett.} \textbf{\bibinfo{volume}{98}},
  \bibinfo{eid}{070403} (\bibinfo{year}{2007}).

\bibitem[{\citenamefont{Kakashvili and Bolech}(2009)}]{kakashvili09}
\bibinfo{author}{\bibfnamefont{P.}~\bibnamefont{Kakashvili}} \bibnamefont{and}
  \bibinfo{author}{\bibfnamefont{C.~J.} \bibnamefont{Bolech}},
  \bibinfo{journal}{Phys. Rev. A} \textbf{\bibinfo{volume}{79}},
  \bibinfo{eid}{041603} (\bibinfo{year}{2009}).

\bibitem[{\citenamefont{Yang}(2001)}]{yang01}
\bibinfo{author}{\bibfnamefont{K.}~\bibnamefont{Yang}}, \bibinfo{journal}{Phys.
  Rev. B} \textbf{\bibinfo{volume}{63}}, \bibinfo{pages}{140511}
  (\bibinfo{year}{2001}).

\bibitem[{\citenamefont{Feiguin and Heidrich-Meisner}(2007)}]{feiguin07}
\bibinfo{author}{\bibfnamefont{A.~E.} \bibnamefont{Feiguin}} \bibnamefont{and}
  \bibinfo{author}{\bibfnamefont{F.}~\bibnamefont{Heidrich-Meisner}},
  \bibinfo{journal}{Phys. Rev. B} \textbf{\bibinfo{volume}{76}},
  \bibinfo{pages}{220508(R)} (\bibinfo{year}{2007}).

\bibitem[{\citenamefont{Batrouni et~al.}(2008)\citenamefont{Batrouni, Huntley,
  Rousseau, and Scalettar}}]{batrouni08}
\bibinfo{author}{\bibfnamefont{G.~G.} \bibnamefont{Batrouni}},
  \bibinfo{author}{\bibfnamefont{M.~H.} \bibnamefont{Huntley}},
  \bibinfo{author}{\bibfnamefont{V.~G.} \bibnamefont{Rousseau}},
  \bibnamefont{and} \bibinfo{author}{\bibfnamefont{R.~T.}
  \bibnamefont{Scalettar}}, \bibinfo{journal}{Phys. Rev. Lett.}
  \textbf{\bibinfo{volume}{100}}, \bibinfo{eid}{116405} (\bibinfo{year}{2008}).

\bibitem[{\citenamefont{Tezuka and Ueda}(2008)}]{tezuka08}
\bibinfo{author}{\bibfnamefont{M.}~\bibnamefont{Tezuka}} \bibnamefont{and}
  \bibinfo{author}{\bibfnamefont{M.}~\bibnamefont{Ueda}},
  \bibinfo{journal}{Phys. Rev. Lett.} \textbf{\bibinfo{volume}{100}},
  \bibinfo{pages}{110403} (\bibinfo{year}{2008}).

\bibitem[{\citenamefont{Casula et~al.}(2008)\citenamefont{Casula, Ceperley, and
  Mueller}}]{casula08}
\bibinfo{author}{\bibfnamefont{M.}~\bibnamefont{Casula}},
  \bibinfo{author}{\bibfnamefont{D.~M.} \bibnamefont{Ceperley}},
  \bibnamefont{and} \bibinfo{author}{\bibfnamefont{E.~J.}
  \bibnamefont{Mueller}}, \bibinfo{journal}{Phys. Rev. A}
  \textbf{\bibinfo{volume}{78}}, \bibinfo{eid}{033607} (\bibinfo{year}{2008}).

\bibitem[{\citenamefont{L\"uscher et~al.}(2008)\citenamefont{L\"uscher, Noack,
  and L\"auchli}}]{luescher08}
\bibinfo{author}{\bibfnamefont{A.}~\bibnamefont{L\"uscher}},
  \bibinfo{author}{\bibfnamefont{R.~M.} \bibnamefont{Noack}}, \bibnamefont{and}
  \bibinfo{author}{\bibfnamefont{A.~M.} \bibnamefont{L\"auchli}},
  \bibinfo{journal}{Phys. Rev. A} \textbf{\bibinfo{volume}{78}},
  \bibinfo{pages}{013637} (\bibinfo{year}{2008}).

\bibitem[{\citenamefont{Bakhtiari et~al.}(2008)\citenamefont{Bakhtiari,
  Leskinen, and T\"{o}rm\"{a}}}]{bakhtiari08}
\bibinfo{author}{\bibfnamefont{M.~R.} \bibnamefont{Bakhtiari}},
  \bibinfo{author}{\bibfnamefont{M.~J.} \bibnamefont{Leskinen}},
  \bibnamefont{and}
  \bibinfo{author}{\bibfnamefont{P.}~\bibnamefont{T\"{o}rm\"{a}}},
  \bibinfo{journal}{Phys. Rev. Lett.} \textbf{\bibinfo{volume}{101}},
  \bibinfo{eid}{120404} (\bibinfo{year}{2008}).

\bibitem[{\citenamefont{Roscilde et~al.}(2009)\citenamefont{Roscilde,
  Rodriguez, Eckert, Romero-Isart, Lewenstein, Polzik, and
  Sanpera}}]{roscilde09}
\bibinfo{author}{\bibfnamefont{T.}~\bibnamefont{Roscilde}},
  \bibinfo{author}{\bibfnamefont{M.}~\bibnamefont{Rodriguez}},
  \bibinfo{author}{\bibfnamefont{K.}~\bibnamefont{Eckert}},
  \bibinfo{author}{\bibfnamefont{O.}~\bibnamefont{Romero-Isart}},
  \bibinfo{author}{\bibfnamefont{M.}~\bibnamefont{Lewenstein}},
  \bibinfo{author}{\bibfnamefont{E.}~\bibnamefont{Polzik}}, \bibnamefont{and}
  \bibinfo{author}{\bibfnamefont{A.}~\bibnamefont{Sanpera}},
  \bibinfo{journal}{New. J. Phys.} \textbf{\bibinfo{volume}{11}},
  \bibinfo{pages}{055041} (\bibinfo{year}{2009}).

\bibitem[{\citenamefont{Edge and Cooper}(2009)}]{edge09}
\bibinfo{author}{\bibfnamefont{J.~M.} \bibnamefont{Edge}} \bibnamefont{and}
  \bibinfo{author}{\bibfnamefont{N.~R.} \bibnamefont{Cooper}},
  \bibinfo{journal}{Phys. Rev. Lett.} \textbf{\bibinfo{volume}{103}},
  \bibinfo{pages}{065301} (\bibinfo{year}{2009}).

\bibitem[{\citenamefont{Kajala et~al.}(2011{\natexlab{a}})\citenamefont{Kajala,
  Massel, and T\"orm\"a}}]{kajala11a}
\bibinfo{author}{\bibfnamefont{J.}~\bibnamefont{Kajala}},
  \bibinfo{author}{\bibfnamefont{F.}~\bibnamefont{Massel}}, \bibnamefont{and}
  \bibinfo{author}{\bibfnamefont{P.}~\bibnamefont{T\"orm\"a}},
  \bibinfo{journal}{Phys. Rev. A} \textbf{\bibinfo{volume}{84}},
  \bibinfo{pages}{041601} (\bibinfo{year}{2011}{\natexlab{a}}).

\bibitem[{\citenamefont{Bolech et~al.}(2012)\citenamefont{Bolech,
  Heidrich-Meisner, Langer, McCulloch, Orso, and Rigol}}]{bolech12}
\bibinfo{author}{\bibfnamefont{C.~J.} \bibnamefont{Bolech}},
  \bibinfo{author}{\bibfnamefont{F.}~\bibnamefont{Heidrich-Meisner}},
  \bibinfo{author}{\bibfnamefont{S.}~\bibnamefont{Langer}},
  \bibinfo{author}{\bibfnamefont{I.~P.} \bibnamefont{McCulloch}},
  \bibinfo{author}{\bibfnamefont{G.}~\bibnamefont{Orso}}, \bibnamefont{and}
  \bibinfo{author}{\bibfnamefont{M.}~\bibnamefont{Rigol}},
  \bibinfo{journal}{Phys. Rev. Lett.} \textbf{\bibinfo{volume}{109}},
  \bibinfo{pages}{110602} (\bibinfo{year}{2012}).

\bibitem[{\citenamefont{Lu et~al.}(2012)\citenamefont{Lu, Baksmaty, Bolech, and
  Pu}}]{lu12}
\bibinfo{author}{\bibfnamefont{H.}~\bibnamefont{Lu}},
  \bibinfo{author}{\bibfnamefont{L.~O.} \bibnamefont{Baksmaty}},
  \bibinfo{author}{\bibfnamefont{C.~J.} \bibnamefont{Bolech}},
  \bibnamefont{and} \bibinfo{author}{\bibfnamefont{H.}~\bibnamefont{Pu}},
  \bibinfo{journal}{Phys. Rev. Lett.} \textbf{\bibinfo{volume}{108}},
  \bibinfo{pages}{225302} (\bibinfo{year}{2012}).

\bibitem[{\citenamefont{Schneider et~al.}(2008)\citenamefont{Schneider,
  Hackerm\"uller, Will, Best, Bloch, Costi, Helmes, Rasch, and
  Rosch}}]{schneider08}
\bibinfo{author}{\bibfnamefont{U.}~\bibnamefont{Schneider}},
  \bibinfo{author}{\bibfnamefont{L.}~\bibnamefont{Hackerm\"uller}},
  \bibinfo{author}{\bibfnamefont{S.}~\bibnamefont{Will}},
  \bibinfo{author}{\bibfnamefont{T.}~\bibnamefont{Best}},
  \bibinfo{author}{\bibfnamefont{I.}~\bibnamefont{Bloch}},
  \bibinfo{author}{\bibfnamefont{T.~A.} \bibnamefont{Costi}},
  \bibinfo{author}{\bibfnamefont{R.~W.} \bibnamefont{Helmes}},
  \bibinfo{author}{\bibfnamefont{D.}~\bibnamefont{Rasch}}, \bibnamefont{and}
  \bibinfo{author}{\bibfnamefont{A.}~\bibnamefont{Rosch}},
  \bibinfo{journal}{Science} \textbf{\bibinfo{volume}{322}},
  \bibinfo{pages}{1520} (\bibinfo{year}{2008}).

\bibitem[{\citenamefont{J\"ordens et~al.}(2008)\citenamefont{J\"ordens,
  Strohmaier, G\"unter, Moritz, and Esslinger}}]{joerdens08}
\bibinfo{author}{\bibfnamefont{R.}~\bibnamefont{J\"ordens}},
  \bibinfo{author}{\bibfnamefont{N.}~\bibnamefont{Strohmaier}},
  \bibinfo{author}{\bibfnamefont{K.}~\bibnamefont{G\"unter}},
  \bibinfo{author}{\bibfnamefont{H.}~\bibnamefont{Moritz}}, \bibnamefont{and}
  \bibinfo{author}{\bibfnamefont{T.}~\bibnamefont{Esslinger}},
  \bibinfo{journal}{Nature (London)} \textbf{\bibinfo{volume}{455}},
  \bibinfo{pages}{204} (\bibinfo{year}{2008}).

\bibitem[{\citenamefont{Hart et~al.}(2015)\citenamefont{Hart, Duarte, Yang,
  Liu, Paiva, Khatami, Scalettar, Trivedi, Huse, and Hulet}}]{hart14}
\bibinfo{author}{\bibfnamefont{R.~A.} \bibnamefont{Hart}},
  \bibinfo{author}{\bibfnamefont{P.~M.} \bibnamefont{Duarte}},
  \bibinfo{author}{\bibfnamefont{T.-L.} \bibnamefont{Yang}},
  \bibinfo{author}{\bibfnamefont{X.}~\bibnamefont{Liu}},
  \bibinfo{author}{\bibfnamefont{T.}~\bibnamefont{Paiva}},
  \bibinfo{author}{\bibfnamefont{E.}~\bibnamefont{Khatami}},
  \bibinfo{author}{\bibfnamefont{R.~T.} \bibnamefont{Scalettar}},
  \bibinfo{author}{\bibfnamefont{N.}~\bibnamefont{Trivedi}},
  \bibinfo{author}{\bibfnamefont{D.~A.} \bibnamefont{Huse}}, \bibnamefont{and}
  \bibinfo{author}{\bibfnamefont{R.~G.} \bibnamefont{Hulet}},
  \bibinfo{journal}{Nature} \textbf{\bibinfo{volume}{519}},
  \bibinfo{pages}{211} (\bibinfo{year}{2015}).

\bibitem[{\citenamefont{Kinoshita et~al.}(2006)\citenamefont{Kinoshita, Wenger,
  and D.~Weiss}}]{kinoshita06}
\bibinfo{author}{\bibfnamefont{T.}~\bibnamefont{Kinoshita}},
  \bibinfo{author}{\bibfnamefont{T.}~\bibnamefont{Wenger}}, \bibnamefont{and}
  \bibinfo{author}{\bibfnamefont{S.}~\bibnamefont{D.~Weiss}},
  \bibinfo{journal}{Nature (London)} \textbf{\bibinfo{volume}{440}},
  \bibinfo{pages}{900} (\bibinfo{year}{2006}).

\bibitem[{\citenamefont{Hofferberth et~al.}(2007)\citenamefont{Hofferberth,
  Lesanovsky, Fischer, Schumm, and Schmiedmayer}}]{hofferberth07}
\bibinfo{author}{\bibfnamefont{S.}~\bibnamefont{Hofferberth}},
  \bibinfo{author}{\bibfnamefont{I.}~\bibnamefont{Lesanovsky}},
  \bibinfo{author}{\bibfnamefont{B.}~\bibnamefont{Fischer}},
  \bibinfo{author}{\bibfnamefont{T.}~\bibnamefont{Schumm}}, \bibnamefont{and}
  \bibinfo{author}{\bibfnamefont{J.}~\bibnamefont{Schmiedmayer}},
  \bibinfo{journal}{Nature (London)} \textbf{\bibinfo{volume}{449}},
  \bibinfo{pages}{324} (\bibinfo{year}{2007}).

\bibitem[{\citenamefont{Kasztelan et~al.}(2011)\citenamefont{Kasztelan,
  Trotzky, Chen, Bloch, McCulloch, Schollw\"ock, and Orso}}]{kasztelan12}
\bibinfo{author}{\bibfnamefont{C.}~\bibnamefont{Kasztelan}},
  \bibinfo{author}{\bibfnamefont{S.}~\bibnamefont{Trotzky}},
  \bibinfo{author}{\bibfnamefont{Y.-A.} \bibnamefont{Chen}},
  \bibinfo{author}{\bibfnamefont{I.}~\bibnamefont{Bloch}},
  \bibinfo{author}{\bibfnamefont{I.~P.} \bibnamefont{McCulloch}},
  \bibinfo{author}{\bibfnamefont{U.}~\bibnamefont{Schollw\"ock}},
  \bibnamefont{and} \bibinfo{author}{\bibfnamefont{G.}~\bibnamefont{Orso}},
  \bibinfo{journal}{Phys. Rev. Lett.} \textbf{\bibinfo{volume}{106}},
  \bibinfo{pages}{155302} (\bibinfo{year}{2011}).

\bibitem[{\citenamefont{Trotzky et~al.}(2012)\citenamefont{Trotzky, Chen,
  Flesch, McCulloch, Schollw\"ock, Eisert, and Bloch}}]{trotzky12}
\bibinfo{author}{\bibfnamefont{S.}~\bibnamefont{Trotzky}},
  \bibinfo{author}{\bibfnamefont{Y.-A.} \bibnamefont{Chen}},
  \bibinfo{author}{\bibfnamefont{A.}~\bibnamefont{Flesch}},
  \bibinfo{author}{\bibfnamefont{I.~P.} \bibnamefont{McCulloch}},
  \bibinfo{author}{\bibfnamefont{U.}~\bibnamefont{Schollw\"ock}},
  \bibinfo{author}{\bibfnamefont{J.}~\bibnamefont{Eisert}}, \bibnamefont{and}
  \bibinfo{author}{\bibfnamefont{I.}~\bibnamefont{Bloch}},
  \bibinfo{journal}{Nature Phys.} \textbf{\bibinfo{volume}{8}},
  \bibinfo{pages}{325} (\bibinfo{year}{2012}).

\bibitem[{\citenamefont{Gring et~al.}(2012)\citenamefont{Gring, Kuhnert,
  Langen, Kitagawa, Rauer, Schreitl, Mazets, Smith, Demler, and
  Schmiedmayer}}]{gring12}
\bibinfo{author}{\bibfnamefont{M.}~\bibnamefont{Gring}},
  \bibinfo{author}{\bibfnamefont{M.}~\bibnamefont{Kuhnert}},
  \bibinfo{author}{\bibfnamefont{T.}~\bibnamefont{Langen}},
  \bibinfo{author}{\bibfnamefont{T.}~\bibnamefont{Kitagawa}},
  \bibinfo{author}{\bibfnamefont{B.}~\bibnamefont{Rauer}},
  \bibinfo{author}{\bibfnamefont{M.}~\bibnamefont{Schreitl}},
  \bibinfo{author}{\bibfnamefont{I.}~\bibnamefont{Mazets}},
  \bibinfo{author}{\bibfnamefont{D.~A.} \bibnamefont{Smith}},
  \bibinfo{author}{\bibfnamefont{E.}~\bibnamefont{Demler}}, \bibnamefont{and}
  \bibinfo{author}{\bibfnamefont{J.}~\bibnamefont{Schmiedmayer}},
  \bibinfo{journal}{Science} \textbf{\bibinfo{volume}{337}},
  \bibinfo{pages}{6100} (\bibinfo{year}{2012}).

\bibitem[{\citenamefont{Cheneau et~al.}(2012)\citenamefont{Cheneau, Barmettler,
  Poletti, Endres, Schau\ss, Fukuhara, Gross, Bloch, Kollath, and
  Kuhr}}]{cheneau12}
\bibinfo{author}{\bibfnamefont{M.}~\bibnamefont{Cheneau}},
  \bibinfo{author}{\bibfnamefont{P.}~\bibnamefont{Barmettler}},
  \bibinfo{author}{\bibfnamefont{D.}~\bibnamefont{Poletti}},
  \bibinfo{author}{\bibfnamefont{M.}~\bibnamefont{Endres}},
  \bibinfo{author}{\bibfnamefont{P.}~\bibnamefont{Schau\ss}},
  \bibinfo{author}{\bibfnamefont{T.}~\bibnamefont{Fukuhara}},
  \bibinfo{author}{\bibfnamefont{C.}~\bibnamefont{Gross}},
  \bibinfo{author}{\bibfnamefont{I.}~\bibnamefont{Bloch}},
  \bibinfo{author}{\bibfnamefont{C.}~\bibnamefont{Kollath}}, \bibnamefont{and}
  \bibinfo{author}{\bibfnamefont{S.}~\bibnamefont{Kuhr}},
  \bibinfo{journal}{Nature (London)} \textbf{\bibinfo{volume}{481}},
  \bibinfo{pages}{484} (\bibinfo{year}{2012}).

\bibitem[{\citenamefont{Ronzheimer et~al.}(2013)\citenamefont{Ronzheimer,
  Schreiber, Braun, Hodgman, Langer, McCulloch, Heidrich-Meisner, Bloch, and
  Schneider}}]{ronzheimer13}
\bibinfo{author}{\bibfnamefont{J.~P.} \bibnamefont{Ronzheimer}},
  \bibinfo{author}{\bibfnamefont{M.}~\bibnamefont{Schreiber}},
  \bibinfo{author}{\bibfnamefont{S.}~\bibnamefont{Braun}},
  \bibinfo{author}{\bibfnamefont{S.~S.} \bibnamefont{Hodgman}},
  \bibinfo{author}{\bibfnamefont{S.}~\bibnamefont{Langer}},
  \bibinfo{author}{\bibfnamefont{I.~P.} \bibnamefont{McCulloch}},
  \bibinfo{author}{\bibfnamefont{F.}~\bibnamefont{Heidrich-Meisner}},
  \bibinfo{author}{\bibfnamefont{I.}~\bibnamefont{Bloch}}, \bibnamefont{and}
  \bibinfo{author}{\bibfnamefont{U.}~\bibnamefont{Schneider}},
  \bibinfo{journal}{Phys. Rev. Lett.} \textbf{\bibinfo{volume}{110}},
  \bibinfo{pages}{205301} (\bibinfo{year}{2013}).

\bibitem[{\citenamefont{Langen et~al.}(2013)\citenamefont{Langen, Geiger,
  Kuhnert, Rauer, and Schmiedmayer}}]{langen13}
\bibinfo{author}{\bibfnamefont{T.}~\bibnamefont{Langen}},
  \bibinfo{author}{\bibfnamefont{R.}~\bibnamefont{Geiger}},
  \bibinfo{author}{\bibfnamefont{M.}~\bibnamefont{Kuhnert}},
  \bibinfo{author}{\bibfnamefont{B.}~\bibnamefont{Rauer}}, \bibnamefont{and}
  \bibinfo{author}{\bibfnamefont{J.}~\bibnamefont{Schmiedmayer}},
  \bibinfo{journal}{Nature Phys.} \textbf{\bibinfo{volume}{9}},
  \bibinfo{pages}{640} (\bibinfo{year}{2013}).

\bibitem[{\citenamefont{Rigol et~al.}(2008)\citenamefont{Rigol, Dunjko, and
  Olshanii}}]{rigol08}
\bibinfo{author}{\bibfnamefont{M.}~\bibnamefont{Rigol}},
  \bibinfo{author}{\bibfnamefont{V.}~\bibnamefont{Dunjko}}, \bibnamefont{and}
  \bibinfo{author}{\bibfnamefont{M.}~\bibnamefont{Olshanii}},
  \bibinfo{journal}{Nature (London)} \textbf{\bibinfo{volume}{452}},
  \bibinfo{pages}{854} (\bibinfo{year}{2008}).

\bibitem[{\citenamefont{Polkovnikov et~al.}(2011)\citenamefont{Polkovnikov,
  Sengupta, Silva, and Vengalattore}}]{polkovnikov11}
\bibinfo{author}{\bibfnamefont{A.}~\bibnamefont{Polkovnikov}},
  \bibinfo{author}{\bibfnamefont{K.}~\bibnamefont{Sengupta}},
  \bibinfo{author}{\bibfnamefont{A.}~\bibnamefont{Silva}}, \bibnamefont{and}
  \bibinfo{author}{\bibfnamefont{M.}~\bibnamefont{Vengalattore}},
  \bibinfo{journal}{Rev. Mod. Phys} \textbf{\bibinfo{volume}{83}},
  \bibinfo{pages}{863} (\bibinfo{year}{2011}).

\bibitem[{\citenamefont{Eisert et~al.}(2015)\citenamefont{Eisert, Friesdorf,
  and Gogolin}}]{eisert14}
\bibinfo{author}{\bibfnamefont{J.}~\bibnamefont{Eisert}},
  \bibinfo{author}{\bibfnamefont{M.}~\bibnamefont{Friesdorf}},
  \bibnamefont{and} \bibinfo{author}{\bibfnamefont{C.}~\bibnamefont{Gogolin}},
  \bibinfo{journal}{Nature Phys.} \textbf{\bibinfo{volume}{11}},
  \bibinfo{pages}{124} (\bibinfo{year}{2015}).

\bibitem[{\citenamefont{Langen et~al.}(unpublished)\citenamefont{Langen,
  Geiger, and Schmiedmayer}}]{langen14}
\bibinfo{author}{\bibfnamefont{T.}~\bibnamefont{Langen}},
  \bibinfo{author}{\bibfnamefont{R.}~\bibnamefont{Geiger}}, \bibnamefont{and}
  \bibinfo{author}{\bibfnamefont{J.}~\bibnamefont{Schmiedmayer}}, p.
  \bibinfo{pages}{arXiv:1408.6377} (\bibinfo{year}{unpublished}).

\bibitem[{\citenamefont{Rigol et~al.}(2007)\citenamefont{Rigol, Dunjko,
  Yurovsky, and Olshanii}}]{rigol07}
\bibinfo{author}{\bibfnamefont{M.}~\bibnamefont{Rigol}},
  \bibinfo{author}{\bibfnamefont{V.}~\bibnamefont{Dunjko}},
  \bibinfo{author}{\bibfnamefont{V.}~\bibnamefont{Yurovsky}}, \bibnamefont{and}
  \bibinfo{author}{\bibfnamefont{M.}~\bibnamefont{Olshanii}},
  \bibinfo{journal}{Phys. Rev. Lett.} \textbf{\bibinfo{volume}{98}},
  \bibinfo{pages}{050405} (\bibinfo{year}{2007}).

\bibitem[{\citenamefont{Essler et~al.}(2012)\citenamefont{Essler, Evangelisti,
  and Fagotti}}]{essler12}
\bibinfo{author}{\bibfnamefont{F.~H.~L.} \bibnamefont{Essler}},
  \bibinfo{author}{\bibfnamefont{S.}~\bibnamefont{Evangelisti}},
  \bibnamefont{and} \bibinfo{author}{\bibfnamefont{M.}~\bibnamefont{Fagotti}},
  \bibinfo{journal}{Phys. Rev. Lett.} \textbf{\bibinfo{volume}{109}},
  \bibinfo{pages}{247206} (\bibinfo{year}{2012}).

\bibitem[{\citenamefont{Caux and Konik}(2012)}]{caux12}
\bibinfo{author}{\bibfnamefont{J.-S.} \bibnamefont{Caux}} \bibnamefont{and}
  \bibinfo{author}{\bibfnamefont{R.~M.} \bibnamefont{Konik}},
  \bibinfo{journal}{Phys. Rev. Lett.} \textbf{\bibinfo{volume}{109}},
  \bibinfo{pages}{175301} (\bibinfo{year}{2012}).

\bibitem[{\citenamefont{Mierzejewski et~al.}(2014)\citenamefont{Mierzejewski,
  Prelov\ifmmode~\check{s}\else \v{s}\fi{}ek, and Prosen}}]{Mierzejewski14}
\bibinfo{author}{\bibfnamefont{M.}~\bibnamefont{Mierzejewski}},
  \bibinfo{author}{\bibfnamefont{P.}~\bibnamefont{Prelov\ifmmode~\check{s}\else
  \v{s}\fi{}ek}}, \bibnamefont{and}
  \bibinfo{author}{\bibfnamefont{T.}~\bibnamefont{Prosen}},
  \bibinfo{journal}{Phys. Rev. Lett.} \textbf{\bibinfo{volume}{113}},
  \bibinfo{pages}{020602} (\bibinfo{year}{2014}).

\bibitem[{\citenamefont{Wouters et~al.}(2014)\citenamefont{Wouters, De~Nardis,
  Brockmann, Fioretto, Rigol, and Caux}}]{wouters14}
\bibinfo{author}{\bibfnamefont{B.}~\bibnamefont{Wouters}},
  \bibinfo{author}{\bibfnamefont{J.}~\bibnamefont{De~Nardis}},
  \bibinfo{author}{\bibfnamefont{M.}~\bibnamefont{Brockmann}},
  \bibinfo{author}{\bibfnamefont{D.}~\bibnamefont{Fioretto}},
  \bibinfo{author}{\bibfnamefont{M.}~\bibnamefont{Rigol}}, \bibnamefont{and}
  \bibinfo{author}{\bibfnamefont{J.-S.} \bibnamefont{Caux}},
  \bibinfo{journal}{Phys. Rev. Lett.} \textbf{\bibinfo{volume}{113}},
  \bibinfo{pages}{117202} (\bibinfo{year}{2014}).

\bibitem[{\citenamefont{Pozsgay et~al.}(2014)\citenamefont{Pozsgay, Mesty\'an,
  Werner, Kormos, Zar\'and, and Tak\'acs}}]{pozsgay14}
\bibinfo{author}{\bibfnamefont{B.}~\bibnamefont{Pozsgay}},
  \bibinfo{author}{\bibfnamefont{M.}~\bibnamefont{Mesty\'an}},
  \bibinfo{author}{\bibfnamefont{M.~A.} \bibnamefont{Werner}},
  \bibinfo{author}{\bibfnamefont{M.}~\bibnamefont{Kormos}},
  \bibinfo{author}{\bibfnamefont{G.}~\bibnamefont{Zar\'and}}, \bibnamefont{and}
  \bibinfo{author}{\bibfnamefont{G.}~\bibnamefont{Tak\'acs}},
  \bibinfo{journal}{Phys. Rev. Lett.} \textbf{\bibinfo{volume}{113}},
  \bibinfo{pages}{117203} (\bibinfo{year}{2014}).

\bibitem[{\citenamefont{Goldstein and Andrei}(2014)}]{goldstein14}
\bibinfo{author}{\bibfnamefont{G.}~\bibnamefont{Goldstein}} \bibnamefont{and}
  \bibinfo{author}{\bibfnamefont{N.}~\bibnamefont{Andrei}},
  \bibinfo{journal}{Phys. Rev. A} \textbf{\bibinfo{volume}{90}},
  \bibinfo{pages}{043625} (\bibinfo{year}{2014}).

\bibitem[{\citenamefont{Alba}(unpublished)}]{alba14}
\bibinfo{author}{\bibfnamefont{V.}~\bibnamefont{Alba}}, p.
  \bibinfo{pages}{arXiv:1409.6096} (\bibinfo{year}{unpublished}).

\bibitem[{\citenamefont{Strohmaier et~al.}(2010)\citenamefont{Strohmaier,
  Greif, J\"ordens, Tarruell, Moritz, Esslinger, Sensarma, Pekker, Altman, and
  Demler}}]{strohmaier10}
\bibinfo{author}{\bibfnamefont{N.}~\bibnamefont{Strohmaier}},
  \bibinfo{author}{\bibfnamefont{D.}~\bibnamefont{Greif}},
  \bibinfo{author}{\bibfnamefont{R.}~\bibnamefont{J\"ordens}},
  \bibinfo{author}{\bibfnamefont{L.}~\bibnamefont{Tarruell}},
  \bibinfo{author}{\bibfnamefont{H.}~\bibnamefont{Moritz}},
  \bibinfo{author}{\bibfnamefont{T.}~\bibnamefont{Esslinger}},
  \bibinfo{author}{\bibfnamefont{R.}~\bibnamefont{Sensarma}},
  \bibinfo{author}{\bibfnamefont{D.}~\bibnamefont{Pekker}},
  \bibinfo{author}{\bibfnamefont{E.}~\bibnamefont{Altman}}, \bibnamefont{and}
  \bibinfo{author}{\bibfnamefont{E.}~\bibnamefont{Demler}},
  \bibinfo{journal}{Phys. Rev. Lett.} \textbf{\bibinfo{volume}{104}},
  \bibinfo{pages}{080401} (\bibinfo{year}{2010}).

\bibitem[{\citenamefont{Schneider et~al.}(2012)\citenamefont{Schneider,
  Hackerm\"uller, Ronzheimer, Will, Braun, Best, Bloch, Demler, Mandt, Rasch
  et~al.}}]{schneider12}
\bibinfo{author}{\bibfnamefont{U.}~\bibnamefont{Schneider}},
  \bibinfo{author}{\bibfnamefont{L.}~\bibnamefont{Hackerm\"uller}},
  \bibinfo{author}{\bibfnamefont{J.~P.} \bibnamefont{Ronzheimer}},
  \bibinfo{author}{\bibfnamefont{S.}~\bibnamefont{Will}},
  \bibinfo{author}{\bibfnamefont{S.}~\bibnamefont{Braun}},
  \bibinfo{author}{\bibfnamefont{T.}~\bibnamefont{Best}},
  \bibinfo{author}{\bibfnamefont{I.}~\bibnamefont{Bloch}},
  \bibinfo{author}{\bibfnamefont{E.}~\bibnamefont{Demler}},
  \bibinfo{author}{\bibfnamefont{S.}~\bibnamefont{Mandt}},
  \bibinfo{author}{\bibfnamefont{D.}~\bibnamefont{Rasch}},
  \bibnamefont{et~al.}, \bibinfo{journal}{Nature Phys.}
  \textbf{\bibinfo{volume}{8}}, \bibinfo{pages}{213} (\bibinfo{year}{2012}).

\bibitem[{\citenamefont{Will et~al.}(2015)\citenamefont{Will, Iyer, and
  Rigol}}]{will14}
\bibinfo{author}{\bibfnamefont{S.}~\bibnamefont{Will}},
  \bibinfo{author}{\bibfnamefont{D.}~\bibnamefont{Iyer}}, \bibnamefont{and}
  \bibinfo{author}{\bibfnamefont{M.}~\bibnamefont{Rigol}},
  \bibinfo{journal}{Nature Commun.} \textbf{\bibinfo{volume}{6}},
  \bibinfo{pages}{6009} (\bibinfo{year}{2015}).

\bibitem[{\citenamefont{Pertot et~al.}(2014)\citenamefont{Pertot, Sheikhan,
  Cocchi, Miller, Bohn, Koschorreck, K\"ohl, and Kollath}}]{perthot14}
\bibinfo{author}{\bibfnamefont{D.}~\bibnamefont{Pertot}},
  \bibinfo{author}{\bibfnamefont{A.}~\bibnamefont{Sheikhan}},
  \bibinfo{author}{\bibfnamefont{E.}~\bibnamefont{Cocchi}},
  \bibinfo{author}{\bibfnamefont{L.~A.} \bibnamefont{Miller}},
  \bibinfo{author}{\bibfnamefont{J.~E.} \bibnamefont{Bohn}},
  \bibinfo{author}{\bibfnamefont{M.}~\bibnamefont{Koschorreck}},
  \bibinfo{author}{\bibfnamefont{M.}~\bibnamefont{K\"ohl}}, \bibnamefont{and}
  \bibinfo{author}{\bibfnamefont{C.}~\bibnamefont{Kollath}},
  \bibinfo{journal}{Phys. Rev. Lett.} \textbf{\bibinfo{volume}{113}},
  \bibinfo{pages}{170403} (\bibinfo{year}{2014}).

\bibitem[{\citenamefont{Manmana et~al.}(2005)\citenamefont{Manmana, Muramatsu,
  and Noack}}]{manmana05}
\bibinfo{author}{\bibfnamefont{S.~R.} \bibnamefont{Manmana}},
  \bibinfo{author}{\bibfnamefont{A.}~\bibnamefont{Muramatsu}},
  \bibnamefont{and} \bibinfo{author}{\bibfnamefont{R.~M.} \bibnamefont{Noack}},
  \bibinfo{journal}{AIP Conf. Proc.} \textbf{\bibinfo{volume}{789}},
  \bibinfo{pages}{269} (\bibinfo{year}{2005}).

\bibitem[{\citenamefont{Sandvik}(2010)}]{sandvik}
\bibinfo{author}{\bibfnamefont{A.}~\bibnamefont{Sandvik}},
  \bibinfo{journal}{AIP Conf. Proc.} \textbf{\bibinfo{volume}{1297}},
  \bibinfo{pages}{135} (\bibinfo{year}{2010}).

\bibitem[{\citenamefont{Vidal}(2004)}]{vidal04}
\bibinfo{author}{\bibfnamefont{G.}~\bibnamefont{Vidal}},
  \bibinfo{journal}{Phys. Rev. Lett.} \textbf{\bibinfo{volume}{93}},
  \bibinfo{pages}{040502} (\bibinfo{year}{2004}).

\bibitem[{\citenamefont{Daley et~al.}(2004)\citenamefont{Daley, Kollath,
  Schollw\"ock, and Vidal}}]{daley04}
\bibinfo{author}{\bibfnamefont{A.}~\bibnamefont{Daley}},
  \bibinfo{author}{\bibfnamefont{C.}~\bibnamefont{Kollath}},
  \bibinfo{author}{\bibfnamefont{U.}~\bibnamefont{Schollw\"ock}},
  \bibnamefont{and} \bibinfo{author}{\bibfnamefont{G.}~\bibnamefont{Vidal}},
  \bibinfo{journal}{J. Stat. Mech.: Theory Exp.}
  \textbf{\bibinfo{volume}{{\rm(2004)}}}, \bibinfo{pages}{P04005}
  (\bibinfo{year}{2004}).

\bibitem[{\citenamefont{White and Feiguin}(2004)}]{white04}
\bibinfo{author}{\bibfnamefont{S.~R.} \bibnamefont{White}} \bibnamefont{and}
  \bibinfo{author}{\bibfnamefont{A.~E.} \bibnamefont{Feiguin}},
  \bibinfo{journal}{Phys. Rev. Lett.} \textbf{\bibinfo{volume}{93}},
  \bibinfo{pages}{076401} (\bibinfo{year}{2004}).

\bibitem[{\citenamefont{Bernier et~al.}(2011)\citenamefont{Bernier, Roux, and
  Kollath}}]{bernier11}
\bibinfo{author}{\bibfnamefont{J.-S.} \bibnamefont{Bernier}},
  \bibinfo{author}{\bibfnamefont{G.}~\bibnamefont{Roux}}, \bibnamefont{and}
  \bibinfo{author}{\bibfnamefont{C.}~\bibnamefont{Kollath}},
  \bibinfo{journal}{Phys. Rev. Lett.} \textbf{\bibinfo{volume}{106}},
  \bibinfo{pages}{200601} (\bibinfo{year}{2011}).

\bibitem[{\citenamefont{Bernier et~al.}(2012)\citenamefont{Bernier, Poletti,
  Barmettler, Roux, and Kollath}}]{bernier12}
\bibinfo{author}{\bibfnamefont{J.-S.} \bibnamefont{Bernier}},
  \bibinfo{author}{\bibfnamefont{D.}~\bibnamefont{Poletti}},
  \bibinfo{author}{\bibfnamefont{P.}~\bibnamefont{Barmettler}},
  \bibinfo{author}{\bibfnamefont{G.}~\bibnamefont{Roux}}, \bibnamefont{and}
  \bibinfo{author}{\bibfnamefont{C.}~\bibnamefont{Kollath}},
  \bibinfo{journal}{Phys. Rev. A} \textbf{\bibinfo{volume}{85}},
  \bibinfo{pages}{033641} (\bibinfo{year}{2012}).

\bibitem[{\citenamefont{Moeckel and Kehrein}(2008)}]{moeckel08}
\bibinfo{author}{\bibfnamefont{M.}~\bibnamefont{Moeckel}} \bibnamefont{and}
  \bibinfo{author}{\bibfnamefont{S.}~\bibnamefont{Kehrein}},
  \bibinfo{journal}{Phys. Rev. Lett.} \textbf{\bibinfo{volume}{100}},
  \bibinfo{eid}{175702} (\bibinfo{year}{2008}).

\bibitem[{\citenamefont{Kollar and Eckstein}(2008)}]{kollar08}
\bibinfo{author}{\bibfnamefont{M.}~\bibnamefont{Kollar}} \bibnamefont{and}
  \bibinfo{author}{\bibfnamefont{M.}~\bibnamefont{Eckstein}},
  \bibinfo{journal}{Phys. Rev. A} \textbf{\bibinfo{volume}{78}},
  \bibinfo{pages}{013626} (\bibinfo{year}{2008}).

\bibitem[{\citenamefont{Eckstein et~al.}(2009)\citenamefont{Eckstein, Kollar,
  and Werner}}]{eckstein09}
\bibinfo{author}{\bibfnamefont{M.}~\bibnamefont{Eckstein}},
  \bibinfo{author}{\bibfnamefont{M.}~\bibnamefont{Kollar}}, \bibnamefont{and}
  \bibinfo{author}{\bibfnamefont{P.}~\bibnamefont{Werner}},
  \bibinfo{journal}{Phys. Rev. Lett.} \textbf{\bibinfo{volume}{103}},
  \bibinfo{pages}{056403} (\bibinfo{year}{2009}).

\bibitem[{\citenamefont{Eckstein et~al.}(2010)\citenamefont{Eckstein, Kollar,
  and Werner}}]{eckstein10}
\bibinfo{author}{\bibfnamefont{M.}~\bibnamefont{Eckstein}},
  \bibinfo{author}{\bibfnamefont{M.}~\bibnamefont{Kollar}}, \bibnamefont{and}
  \bibinfo{author}{\bibfnamefont{P.}~\bibnamefont{Werner}},
  \bibinfo{journal}{Phys. Rev. B} \textbf{\bibinfo{volume}{81}},
  \bibinfo{pages}{115131} (\bibinfo{year}{2010}).

\bibitem[{\citenamefont{Eckstein and Werner}(2011)}]{eckstein11}
\bibinfo{author}{\bibfnamefont{M.}~\bibnamefont{Eckstein}} \bibnamefont{and}
  \bibinfo{author}{\bibfnamefont{P.}~\bibnamefont{Werner}},
  \bibinfo{journal}{Phys. Rev. B} \textbf{\bibinfo{volume}{84}},
  \bibinfo{pages}{035122} (\bibinfo{year}{2011}).

\bibitem[{\citenamefont{Hamerla and Uhrig}(2013)}]{hamerla13}
\bibinfo{author}{\bibfnamefont{S.~A.} \bibnamefont{Hamerla}} \bibnamefont{and}
  \bibinfo{author}{\bibfnamefont{G.~S.} \bibnamefont{Uhrig}},
  \bibinfo{journal}{Phys. Rev. B} \textbf{\bibinfo{volume}{87}},
  \bibinfo{pages}{064304} (\bibinfo{year}{2013}).

\bibitem[{\citenamefont{Hamerla and Uhrig}(2014)}]{hamerla14}
\bibinfo{author}{\bibfnamefont{S.~A.} \bibnamefont{Hamerla}} \bibnamefont{and}
  \bibinfo{author}{\bibfnamefont{G.~S.} \bibnamefont{Uhrig}},
  \bibinfo{journal}{Phys. Rev. B} \textbf{\bibinfo{volume}{89}},
  \bibinfo{pages}{104301} (\bibinfo{year}{2014}).

\bibitem[{\citenamefont{Heidrich-Meisner
  et~al.}(2009)\citenamefont{Heidrich-Meisner, Manmana, Rigol, Muramatsu,
  Feiguin, and Dagotto}}]{hm09}
\bibinfo{author}{\bibfnamefont{F.}~\bibnamefont{Heidrich-Meisner}},
  \bibinfo{author}{\bibfnamefont{S.~R.} \bibnamefont{Manmana}},
  \bibinfo{author}{\bibfnamefont{M.}~\bibnamefont{Rigol}},
  \bibinfo{author}{\bibfnamefont{A.}~\bibnamefont{Muramatsu}},
  \bibinfo{author}{\bibfnamefont{A.~E.} \bibnamefont{Feiguin}},
  \bibnamefont{and} \bibinfo{author}{\bibfnamefont{E.}~\bibnamefont{Dagotto}},
  \bibinfo{journal}{Phys. Rev. A} \textbf{\bibinfo{volume}{80}},
  \bibinfo{pages}{041603} (\bibinfo{year}{2009}).

\bibitem[{\citenamefont{Ke\ss{}ler et~al.}(2013)\citenamefont{Ke\ss{}ler,
  McCulloch, and Marquardt}}]{kessler13}
\bibinfo{author}{\bibfnamefont{S.}~\bibnamefont{Ke\ss{}ler}},
  \bibinfo{author}{\bibfnamefont{I.~P.} \bibnamefont{McCulloch}},
  \bibnamefont{and}
  \bibinfo{author}{\bibfnamefont{F.}~\bibnamefont{Marquardt}},
  \bibinfo{journal}{New J. Phys.} \textbf{\bibinfo{volume}{15}},
  \bibinfo{pages}{053043} (\bibinfo{year}{2013}).

\bibitem[{\citenamefont{Kajala et~al.}(2011{\natexlab{b}})\citenamefont{Kajala,
  Massel, and T\"orm\"a}}]{kajala11}
\bibinfo{author}{\bibfnamefont{J.}~\bibnamefont{Kajala}},
  \bibinfo{author}{\bibfnamefont{F.}~\bibnamefont{Massel}}, \bibnamefont{and}
  \bibinfo{author}{\bibfnamefont{P.}~\bibnamefont{T\"orm\"a}},
  \bibinfo{journal}{Phys. Rev. Lett.} \textbf{\bibinfo{volume}{106}},
  \bibinfo{pages}{206401} (\bibinfo{year}{2011}{\natexlab{b}}).

\bibitem[{\citenamefont{Bloch et~al.}(2008)\citenamefont{Bloch, Dalibard, and
  Zwerger}}]{bloch08}
\bibinfo{author}{\bibfnamefont{I.}~\bibnamefont{Bloch}},
  \bibinfo{author}{\bibfnamefont{J.}~\bibnamefont{Dalibard}}, \bibnamefont{and}
  \bibinfo{author}{\bibfnamefont{W.}~\bibnamefont{Zwerger}},
  \bibinfo{journal}{Rev. Mod. Phys.} \textbf{\bibinfo{volume}{80}},
  \bibinfo{pages}{885} (\bibinfo{year}{2008}).

\bibitem[{\citenamefont{Schollw\"ock}(2011)}]{schollwoeck11}
\bibinfo{author}{\bibfnamefont{U.}~\bibnamefont{Schollw\"ock}},
  \bibinfo{journal}{Ann. Phys. (N.Y.)} \textbf{\bibinfo{volume}{326}},
  \bibinfo{pages}{96} (\bibinfo{year}{2011}).

\bibitem[{\citenamefont{Schollw\"ock}(2005)}]{schollwoeck05}
\bibinfo{author}{\bibfnamefont{U.}~\bibnamefont{Schollw\"ock}},
  \bibinfo{journal}{Rev. Mod. Phys.} \textbf{\bibinfo{volume}{77}},
  \bibinfo{pages}{259} (\bibinfo{year}{2005}).

\bibitem[{\citenamefont{Sorg et~al.}(2014)\citenamefont{Sorg, Vidmar, Pollet,
  and Heidrich-Meisner}}]{sorg14}
\bibinfo{author}{\bibfnamefont{S.}~\bibnamefont{Sorg}},
  \bibinfo{author}{\bibfnamefont{L.}~\bibnamefont{Vidmar}},
  \bibinfo{author}{\bibfnamefont{L.}~\bibnamefont{Pollet}}, \bibnamefont{and}
  \bibinfo{author}{\bibfnamefont{F.}~\bibnamefont{Heidrich-Meisner}},
  \bibinfo{journal}{Phys. Rev. A} \textbf{\bibinfo{volume}{90}},
  \bibinfo{pages}{033606} (\bibinfo{year}{2014}).

\bibitem[{\citenamefont{Greiner et~al.}(2002)\citenamefont{Greiner, Mandel,
  H\"ansch, and Bloch}}]{greiner02}
\bibinfo{author}{\bibfnamefont{M.}~\bibnamefont{Greiner}},
  \bibinfo{author}{\bibfnamefont{O.}~\bibnamefont{Mandel}},
  \bibinfo{author}{\bibfnamefont{T.}~\bibnamefont{H\"ansch}}, \bibnamefont{and}
  \bibinfo{author}{\bibfnamefont{I.}~\bibnamefont{Bloch}},
  \bibinfo{journal}{Nature (London)} \textbf{\bibinfo{volume}{419}},
  \bibinfo{pages}{51} (\bibinfo{year}{2002}).

\bibitem[{\citenamefont{Iyer et~al.}(2014)\citenamefont{Iyer, Mondaini, Will,
  and Rigol}}]{iyer14}
\bibinfo{author}{\bibfnamefont{D.}~\bibnamefont{Iyer}},
  \bibinfo{author}{\bibfnamefont{R.}~\bibnamefont{Mondaini}},
  \bibinfo{author}{\bibfnamefont{S.}~\bibnamefont{Will}}, \bibnamefont{and}
  \bibinfo{author}{\bibfnamefont{M.}~\bibnamefont{Rigol}},
  \bibinfo{journal}{Phys. Rev. A} \textbf{\bibinfo{volume}{90}},
  \bibinfo{pages}{031602} (\bibinfo{year}{2014}).

\bibitem[{\citenamefont{Iyer et~al.}(2013)\citenamefont{Iyer, Guan, and
  Andrei}}]{iyer13}
\bibinfo{author}{\bibfnamefont{D.}~\bibnamefont{Iyer}},
  \bibinfo{author}{\bibfnamefont{H.}~\bibnamefont{Guan}}, \bibnamefont{and}
  \bibinfo{author}{\bibfnamefont{N.}~\bibnamefont{Andrei}},
  \bibinfo{journal}{Phys. Rev. A} \textbf{\bibinfo{volume}{87}},
  \bibinfo{pages}{053628} (\bibinfo{year}{2013}).

\bibitem[{\citenamefont{Winkler et~al.}(2006)\citenamefont{Winkler, Thalhammer,
  Lang, Grimm, Denschlag, Daley, Kantian, Bahler, and Zoller}}]{winkler06}
\bibinfo{author}{\bibfnamefont{K.}~\bibnamefont{Winkler}},
  \bibinfo{author}{\bibfnamefont{G.}~\bibnamefont{Thalhammer}},
  \bibinfo{author}{\bibfnamefont{F.}~\bibnamefont{Lang}},
  \bibinfo{author}{\bibfnamefont{R.}~\bibnamefont{Grimm}},
  \bibinfo{author}{\bibfnamefont{J.~H.} \bibnamefont{Denschlag}},
  \bibinfo{author}{\bibfnamefont{A.~J.} \bibnamefont{Daley}},
  \bibinfo{author}{\bibfnamefont{A.}~\bibnamefont{Kantian}},
  \bibinfo{author}{\bibfnamefont{H.~P.} \bibnamefont{Bahler}},
  \bibnamefont{and} \bibinfo{author}{\bibfnamefont{P.}~\bibnamefont{Zoller}},
  \bibinfo{journal}{Nature} \textbf{\bibinfo{volume}{441}},
  \bibinfo{pages}{853} (\bibinfo{year}{2006}).

\bibitem[{\citenamefont{Rosch et~al.}(2008)\citenamefont{Rosch, Rasch, Binz,
  and Vojta}}]{rosch08}
\bibinfo{author}{\bibfnamefont{A.}~\bibnamefont{Rosch}},
  \bibinfo{author}{\bibfnamefont{D.}~\bibnamefont{Rasch}},
  \bibinfo{author}{\bibfnamefont{B.}~\bibnamefont{Binz}}, \bibnamefont{and}
  \bibinfo{author}{\bibfnamefont{M.}~\bibnamefont{Vojta}},
  \bibinfo{journal}{Phys. Rev. Lett.} \textbf{\bibinfo{volume}{101}},
  \bibinfo{pages}{265301} (\bibinfo{year}{2008}).

\bibitem[{\citenamefont{Muth et~al.}(2012)\citenamefont{Muth, Petrosyan, and
  Fleischhauer}}]{muth12}
\bibinfo{author}{\bibfnamefont{D.}~\bibnamefont{Muth}},
  \bibinfo{author}{\bibfnamefont{D.}~\bibnamefont{Petrosyan}},
  \bibnamefont{and}
  \bibinfo{author}{\bibfnamefont{M.}~\bibnamefont{Fleischhauer}},
  \bibinfo{journal}{Phys. Rev. A} \textbf{\bibinfo{volume}{85}},
  \bibinfo{pages}{013615} (\bibinfo{year}{2012}).

\bibitem[{\citenamefont{Xia et~al.}(2015)\citenamefont{Xia, Zundel,
  Carrasquilla, Reinhard, Wilson, Rigol, and Weiss}}]{weiss}
\bibinfo{author}{\bibfnamefont{L.}~\bibnamefont{Xia}},
  \bibinfo{author}{\bibfnamefont{L.~A.} \bibnamefont{Zundel}},
  \bibinfo{author}{\bibfnamefont{J.}~\bibnamefont{Carrasquilla}},
  \bibinfo{author}{\bibfnamefont{A.}~\bibnamefont{Reinhard}},
  \bibinfo{author}{\bibfnamefont{J.~M.} \bibnamefont{Wilson}},
  \bibinfo{author}{\bibfnamefont{M.}~\bibnamefont{Rigol}}, \bibnamefont{and}
  \bibinfo{author}{\bibfnamefont{D.~S.} \bibnamefont{Weiss}},
  \bibinfo{journal}{Nature Phys.} \textbf{\bibinfo{volume}{11}},
  \bibinfo{pages}{316} (\bibinfo{year}{2015}).

\bibitem[{\citenamefont{Al-Hassanieh et~al.}(2008)\citenamefont{Al-Hassanieh,
  Reboredo, Feiguin, Gonz\'alez, and Dagotto}}]{alhassanieh08}
\bibinfo{author}{\bibfnamefont{K.~A.} \bibnamefont{Al-Hassanieh}},
  \bibinfo{author}{\bibfnamefont{F.~A.} \bibnamefont{Reboredo}},
  \bibinfo{author}{\bibfnamefont{A.~E.} \bibnamefont{Feiguin}},
  \bibinfo{author}{\bibfnamefont{I.}~\bibnamefont{Gonz\'alez}},
  \bibnamefont{and} \bibinfo{author}{\bibfnamefont{E.}~\bibnamefont{Dagotto}},
  \bibinfo{journal}{Phys. Rev. Lett.} \textbf{\bibinfo{volume}{100}},
  \bibinfo{pages}{166403} (\bibinfo{year}{2008}).

\bibitem[{\citenamefont{Kollath et~al.}(2007)\citenamefont{Kollath, L\"auchli,
  and Altman}}]{kollath07}
\bibinfo{author}{\bibfnamefont{C.}~\bibnamefont{Kollath}},
  \bibinfo{author}{\bibfnamefont{A.~M.} \bibnamefont{L\"auchli}},
  \bibnamefont{and} \bibinfo{author}{\bibfnamefont{E.}~\bibnamefont{Altman}},
  \bibinfo{journal}{Phys. Rev. Lett.} \textbf{\bibinfo{volume}{98}},
  \bibinfo{pages}{180601} (\bibinfo{year}{2007}).

\bibitem[{\citenamefont{Yang}(1989)}]{yang}
\bibinfo{author}{\bibfnamefont{C.~N.} \bibnamefont{Yang}},
  \bibinfo{journal}{Phys. Rev. Lett.} \textbf{\bibinfo{volume}{63}},
  \bibinfo{pages}{2144} (\bibinfo{year}{1989}).

\bibitem[{\citenamefont{Kantian et~al.}(2010)\citenamefont{Kantian, Daley, and
  Zoller}}]{kantian10}
\bibinfo{author}{\bibfnamefont{A.}~\bibnamefont{Kantian}},
  \bibinfo{author}{\bibfnamefont{A.~J.} \bibnamefont{Daley}}, \bibnamefont{and}
  \bibinfo{author}{\bibfnamefont{P.}~\bibnamefont{Zoller}},
  \bibinfo{journal}{Phys. Rev. Lett.} \textbf{\bibinfo{volume}{104}},
  \bibinfo{pages}{240406} (\bibinfo{year}{2010}).

\bibitem[{\citenamefont{Heidrich-Meisner
  et~al.}(2010{\natexlab{a}})\citenamefont{Heidrich-Meisner, Orso, and
  Feiguin}}]{hm10a}
\bibinfo{author}{\bibfnamefont{F.}~\bibnamefont{Heidrich-Meisner}},
  \bibinfo{author}{\bibfnamefont{G.}~\bibnamefont{Orso}}, \bibnamefont{and}
  \bibinfo{author}{\bibfnamefont{A.~E.} \bibnamefont{Feiguin}},
  \bibinfo{journal}{Phys. Rev. A} \textbf{\bibinfo{volume}{81}},
  \bibinfo{pages}{053602} (\bibinfo{year}{2010}{\natexlab{a}}).

\bibitem[{\citenamefont{Tezuka and Ueda}(2010)}]{tezuka10}
\bibinfo{author}{\bibfnamefont{M.}~\bibnamefont{Tezuka}} \bibnamefont{and}
  \bibinfo{author}{\bibfnamefont{M.}~\bibnamefont{Ueda}}, \bibinfo{journal}{New
  J. Phys.} \textbf{\bibinfo{volume}{12}}, \bibinfo{pages}{055029}
  (\bibinfo{year}{2010}).

\bibitem[{\citenamefont{Vidmar et~al.}(2013)\citenamefont{Vidmar, Langer,
  McCulloch, Schneider, Schollw\"ock, and Heidrich-Meisner}}]{vidmar13}
\bibinfo{author}{\bibfnamefont{L.}~\bibnamefont{Vidmar}},
  \bibinfo{author}{\bibfnamefont{S.}~\bibnamefont{Langer}},
  \bibinfo{author}{\bibfnamefont{I.~P.} \bibnamefont{McCulloch}},
  \bibinfo{author}{\bibfnamefont{U.}~\bibnamefont{Schneider}},
  \bibinfo{author}{\bibfnamefont{U.}~\bibnamefont{Schollw\"ock}},
  \bibnamefont{and}
  \bibinfo{author}{\bibfnamefont{F.}~\bibnamefont{Heidrich-Meisner}},
  \bibinfo{journal}{Phys. Rev. B} \textbf{\bibinfo{volume}{88}},
  \bibinfo{pages}{235117} (\bibinfo{year}{2013}).

\bibitem[{\citenamefont{Feiguin and Heidrich-Meisner}(2009)}]{feiguin09}
\bibinfo{author}{\bibfnamefont{A.~E.} \bibnamefont{Feiguin}} \bibnamefont{and}
  \bibinfo{author}{\bibfnamefont{F.}~\bibnamefont{Heidrich-Meisner}},
  \bibinfo{journal}{Phys. Rev. Lett.} \textbf{\bibinfo{volume}{102}},
  \bibinfo{pages}{076403} (\bibinfo{year}{2009}).

\bibitem[{\citenamefont{Foelling et~al.}(2007)\citenamefont{Foelling, Trotzky,
  Cheinet, Feld, Saers, Widera, Mueller, and Bloch}}]{foelling07}
\bibinfo{author}{\bibfnamefont{S.}~\bibnamefont{Foelling}},
  \bibinfo{author}{\bibfnamefont{S.}~\bibnamefont{Trotzky}},
  \bibinfo{author}{\bibfnamefont{P.}~\bibnamefont{Cheinet}},
  \bibinfo{author}{\bibfnamefont{M.}~\bibnamefont{Feld}},
  \bibinfo{author}{\bibfnamefont{R.}~\bibnamefont{Saers}},
  \bibinfo{author}{\bibfnamefont{A.}~\bibnamefont{Widera}},
  \bibinfo{author}{\bibfnamefont{T.}~\bibnamefont{Mueller}}, \bibnamefont{and}
  \bibinfo{author}{\bibfnamefont{I.}~\bibnamefont{Bloch}},
  \bibinfo{journal}{Nature} \textbf{\bibinfo{volume}{448}},
  \bibinfo{pages}{1029} (\bibinfo{year}{2007}).

\bibitem[{\citenamefont{Baur et~al.}(2010)\citenamefont{Baur, Shumway, and
  Mueller}}]{baur10}
\bibinfo{author}{\bibfnamefont{S.~K.} \bibnamefont{Baur}},
  \bibinfo{author}{\bibfnamefont{J.}~\bibnamefont{Shumway}}, \bibnamefont{and}
  \bibinfo{author}{\bibfnamefont{E.~J.} \bibnamefont{Mueller}},
  \bibinfo{journal}{Phys. Rev. A} \textbf{\bibinfo{volume}{81}},
  \bibinfo{pages}{033628} (\bibinfo{year}{2010}).

\bibitem[{\citenamefont{Heidrich-Meisner
  et~al.}(2010{\natexlab{b}})\citenamefont{Heidrich-Meisner, Feiguin,
  Schollw\"ock, and Zwerger}}]{hm10}
\bibinfo{author}{\bibfnamefont{F.}~\bibnamefont{Heidrich-Meisner}},
  \bibinfo{author}{\bibfnamefont{A.~E.} \bibnamefont{Feiguin}},
  \bibinfo{author}{\bibfnamefont{U.}~\bibnamefont{Schollw\"ock}},
  \bibnamefont{and} \bibinfo{author}{\bibfnamefont{W.}~\bibnamefont{Zwerger}},
  \bibinfo{journal}{Phys. Rev. A} \textbf{\bibinfo{volume}{81}},
  \bibinfo{pages}{023629} (\bibinfo{year}{2010}{\natexlab{b}}).

\bibitem[{\citenamefont{Batrouni et~al.}(2009)\citenamefont{Batrouni, Wolak,
  Hebert, and Rousseau}}]{batrouni09}
\bibinfo{author}{\bibfnamefont{G.~G.} \bibnamefont{Batrouni}},
  \bibinfo{author}{\bibfnamefont{M.}~\bibnamefont{Wolak}},
  \bibinfo{author}{\bibfnamefont{F.}~\bibnamefont{Hebert}}, \bibnamefont{and}
  \bibinfo{author}{\bibfnamefont{V.}~\bibnamefont{Rousseau}},
  \bibinfo{journal}{Europhys. Lett.} \textbf{\bibinfo{volume}{86}},
  \bibinfo{pages}{47006} (\bibinfo{year}{2009}).

\bibitem[{\citenamefont{Wang et~al.}(2009)\citenamefont{Wang, Chen, and {Das
  Sarma}}}]{wang09}
\bibinfo{author}{\bibfnamefont{B.}~\bibnamefont{Wang}},
  \bibinfo{author}{\bibfnamefont{H.-D.} \bibnamefont{Chen}}, \bibnamefont{and}
  \bibinfo{author}{\bibfnamefont{S.}~\bibnamefont{{Das Sarma}}},
  \bibinfo{journal}{Phys. Rev. A} \textbf{\bibinfo{volume}{79}},
  \bibinfo{pages}{051604(R)} (\bibinfo{year}{2009}).

\bibitem[{\citenamefont{Orso et~al.}(2010)\citenamefont{Orso, Burovski, and
  Jolicoeur}}]{orso10}
\bibinfo{author}{\bibfnamefont{G.}~\bibnamefont{Orso}},
  \bibinfo{author}{\bibfnamefont{E.}~\bibnamefont{Burovski}}, \bibnamefont{and}
  \bibinfo{author}{\bibfnamefont{T.}~\bibnamefont{Jolicoeur}},
  \bibinfo{journal}{Phys. Rev. Lett.} \textbf{\bibinfo{volume}{104}},
  \bibinfo{pages}{065301} (\bibinfo{year}{2010}).

\bibitem[{\citenamefont{Roux et~al.}(2011)\citenamefont{Roux, Burovski, and
  Jolicoeur}}]{roux11}
\bibinfo{author}{\bibfnamefont{G.}~\bibnamefont{Roux}},
  \bibinfo{author}{\bibfnamefont{E.}~\bibnamefont{Burovski}}, \bibnamefont{and}
  \bibinfo{author}{\bibfnamefont{T.}~\bibnamefont{Jolicoeur}},
  \bibinfo{journal}{Phys. Rev. A} \textbf{\bibinfo{volume}{83}},
  \bibinfo{pages}{053618} (\bibinfo{year}{2011}).

\bibitem[{\citenamefont{Dalmonte et~al.}(2012)\citenamefont{Dalmonte,
  Dieckmann, Roscilde, Hartl, Feiguin, Schollw\"ock, and
  Heidrich-Meisner}}]{dalmonte12}
\bibinfo{author}{\bibfnamefont{M.}~\bibnamefont{Dalmonte}},
  \bibinfo{author}{\bibfnamefont{K.}~\bibnamefont{Dieckmann}},
  \bibinfo{author}{\bibfnamefont{T.}~\bibnamefont{Roscilde}},
  \bibinfo{author}{\bibfnamefont{C.}~\bibnamefont{Hartl}},
  \bibinfo{author}{\bibfnamefont{A.~E.} \bibnamefont{Feiguin}},
  \bibinfo{author}{\bibfnamefont{U.}~\bibnamefont{Schollw\"ock}},
  \bibnamefont{and}
  \bibinfo{author}{\bibfnamefont{F.}~\bibnamefont{Heidrich-Meisner}},
  \bibinfo{journal}{Phys. Rev. A} \textbf{\bibinfo{volume}{85}},
  \bibinfo{pages}{063608} (\bibinfo{year}{2012}).

\bibitem[{\citenamefont{Mandel et~al.}(2003)\citenamefont{Mandel, Greiner,
  Widera, Rom, H\"ansch, and Bloch}}]{mandel}
\bibinfo{author}{\bibfnamefont{O.}~\bibnamefont{Mandel}},
  \bibinfo{author}{\bibfnamefont{M.}~\bibnamefont{Greiner}},
  \bibinfo{author}{\bibfnamefont{A.}~\bibnamefont{Widera}},
  \bibinfo{author}{\bibfnamefont{T.}~\bibnamefont{Rom}},
  \bibinfo{author}{\bibfnamefont{T.~W.} \bibnamefont{H\"ansch}},
  \bibnamefont{and} \bibinfo{author}{\bibfnamefont{I.}~\bibnamefont{Bloch}},
  \bibinfo{journal}{Phys. Rev. Lett.} \textbf{\bibinfo{volume}{91}},
  \bibinfo{pages}{010407} (\bibinfo{year}{2003}).

\bibitem[{\citenamefont{Yao et~al.}(unpublished)\citenamefont{Yao, Laumann,
  Cirac, Lukin, and Moore}}]{moore}
\bibinfo{author}{\bibfnamefont{N.~Y.} \bibnamefont{Yao}},
  \bibinfo{author}{\bibfnamefont{C.~R.} \bibnamefont{Laumann}},
  \bibinfo{author}{\bibfnamefont{J.~I.} \bibnamefont{Cirac}},
  \bibinfo{author}{\bibfnamefont{M.~D.} \bibnamefont{Lukin}}, \bibnamefont{and}
  \bibinfo{author}{\bibfnamefont{J.~E.} \bibnamefont{Moore}}, p.
  \bibinfo{pages}{arXiv:1410.7407} (\bibinfo{year}{unpublished}).

\end{thebibliography}

\end{document}